# Dynamic and programmable cellular-scale granules enable tissue-like materials

Yin Fang[1]†*, Endao Han[1,2]†, Xin-Xing Zhang[1,3,4]†, Yuanwen Jiang[1,3]†, Yiliang Lin[1], Jiuyun Shi[1,3], Jiangbo Wu[1,3], Lingyuan Meng[5], Xiang Gao[1], Philip J. Griffin[5], Xianghui Xiao[6,7], Hsiu-Ming Tsai[8], Hua Zhou[6], Xiaobing Zuo[6], Qing Zhang[6], Miaoqi Chu[6], Qingteng Zhang[6], Ya Gao[6], Leah K. Roth[1,2], Reiner Bleher[9], Zhiyuan Ma[6], Zhang Jiang[6], Jiping Yue[10], Chien-Min Kao[8], Chin-Tu Chen[8], Andrei Tokmakoff[1,3,4], Jin Wang[6], Heinrich M. Jaeger[1,2]*, Bozhi Tian[1,3,4]*

[1]The James Franck Institute, University of Chicago, Chicago, IL 60637.
[2]Department of Physics, University of Chicago, Chicago, IL 60637.
[3]Department of Chemistry, University of Chicago, Chicago, IL 60637.
[4]The Institute for Biophysical Dynamics, University of Chicago, Chicago, IL 60637.
[5]Institute for Molecular Engineering, University of Chicago, Chicago, IL 60637.
[6]Advanced Photon Source, Argonne National Laboratory, Lemont, IL 60439.
[7]National Synchrotron Light Source II, Brookhaven National Laboratory, Upton, NY 11973.
[8]Department of Radiology, University of Chicago, Chicago, IL 60637.
[9]Department of Materials Science and Engineering, Northwestern University, Evanston, IL 60208.
[10]Ben May Department for Cancer Research, University of Chicago, Chicago, IL 60637.
†These authors contributed equally to this work.

*Correspondence to: btian@uchicago.edu, jaeger@uchicago.edu, fangyin123@uchicago.edu

**Abstract:**

Tissue-like materials are required in many robotic systems to improve human-machine interactions. However, the mechanical properties of living tissues are difficult to replicate. Synthetic materials are not usually capable of simultaneously displaying the behaviors of the cellular ensemble and the extracellular matrix. A particular challenge is identification of a cell-like synthetic component which is tightly integrated with its matrix and also responsive to external stimuli at the population level. Here, we demonstrate that cellular-scale hydrated starch granules, an underexplored component in materials science, can turn conventional hydrogels into tissue-like materials when composites are formed. Using several synchrotron-based X-ray techniques, we reveal the mechanically-induced motion and training dynamics of the starch granules in the hydrogel matrix. These dynamic behaviors enable multiple tissue-like properties such as strain-stiffening, anisotropy, mechanical heterogeneity, programmability, mechanochemistry, impact absorption, and self-healability. The starch-hydrogel composites can be processed as robotic skins that maintain these tissue-like characteristics.

**One-sentence Summary**

**Mechanically programmable granular materials in hydrogels enable tissue-like materials for robotic skins.**

**INTRODUCTION**

Soft tissue-like surfaces or interior components are required by many robotic devices to achieve biomimetic functions, enhance robot-human interactions, or reduce invasiveness during medical interventions(*1-5*). Soft materials currently used for robotics are limited to elastomers(*6-8*) (*e.g.*, polydimethylsiloxane) or simple hydrogels(*9, 10*) which usually lack non-equilibrium properties or dynamic behaviors. Tissues such as human skin are multicomponent and hierarchical(*11, 12*), mechanically heterogeneous and anisotropic(*13, 14*), self-healing(*15, 16*), impact absorbing(*17*), and dynamically responsive(*11, 12, 17*). However, these naturally occurring tissue properties are not well recapituated in synthetic materials(*18*) for robotic applications.

In living organisms, extracellular matrices (ECM) provide important biochemical and biomechanical cues for cells. Individual, collective, and emergent cell behavior at the meso- and macro-scale is promoted by ECM networks and drives most physiological processes, including sensory response and feedback, (*19, 20*). Extensive efforts have been made in the biomimetics field to develop ECM-like synthetic polymeric networks(*21-31*); however, an artificial tissue-like material(*1, 32-44*) that can concurrently mimic dynamic cellular-level behavior and ECM-level behavior has yet to be achieved(*45, 46*). The major challenge lies in developing cell-like building blocks that be integrated with ECM-like polymer platforms to display dynamic responses.

Here, we propose that hydrated and cellular-scale granular materials can enable multiscale tissue-like behaviors in synthetic materials. Due to strong intergranular interactions, dense suspensions of granular materials dynamically respond to external stress through rapid phase transformations(*47-49*). When integrated with synthetic hydrogel networks (**Fig. 1A**), the grain-embedded composite may be considered an analog of biological tissue in terms of both structures (hierarchical assembly of cell and ECM) (**fig. S1**) and properties (dynamic responsiveness, collective motion, and pattern formation). To demonstrate proof-of-concept, we use hydrated starch granules (**Figs. 1B-1D**) as the model system because their grain size is similar to the size of single cells. We use a mixture of polyacrylamide (PAA)/alginate (Alg) as the matrix (**Fig. 1D**) because this mixture yields ECM-like viscoelasticity(*29, 50, 51*).

## RESULTS

### Basic building blocks

To prepare the artificial tissue-like construct, we adopted a common thermal-initiator-induced radical polymerization method. Briefly, we vigorously mixed an aqueous suspension of pre-hydrated wheat starch particles with monomers (*i.e.,* acrylamide and Alg) and gelled the entire mixture in a casting mold upon thermal decomposition of the polymerization initiator ammonium persulfate (**table S1**). Unlike the original aqueous suspension of starch granules, the hydrogel composite displays controllable shapes such as membranes and monoliths.

We first studied several static features of the composite. Similar to the ECM structures in living tissues (**fig. S1**), the matrix of the composite shows a nanostructured network. The cellular-scale starch granules have an average grain size of ~11 µm (**Figs. 1B** and **1C**). They are evenly distributed and form tight interfaces with the nanostructured hydrogel network (**Fig. 1D**, **fig. S2**). *In-situ* X-ray scattering of the composite showed that the starch granules are fully hydrated (**fig. S3**). The hydrogel composite exhibited a Young's modulus of ~ 36 kPa, which falls within the range of soft tissues (*i.e.,* 1~500 kPa, **fig. S1**).

### Triggered motility and relaxation dynamics of starch granules

After setting up the starch-in-hydrogel (*i.e.*, a synthetic analog of cell-in-ECM) system, we studied how individual starch granules behave under external stimuli. We fabricated blocks of the hydrogel composite and performed *in situ* X-ray micro-computed tomography (CT) to directly image the equilibrium three-dimensional (3D) arrangement of starch granules under strain (**Fig. 2A**). In aqueous starch suspension, granular packing plays a crucial role in determining the viscosity and shear modulus (when jammed)(*47-49, 52*). In the hybrid gel, the starch loading content was expected to be critical in determining the hydrogel mechanics. In addition to real space images derived from the reconstructed datasets from a hybrid gel (**Fig. 2B** and **fig. S4**), we extracted the spatial distributions of starch granules in the reciprocal space (*i.e., k*-space) using a 3D fast Fourier transform (FFT). We presented the two-dimensional (2D) FFT patterns by averaging the 3D results azimuthally around the stretching direction *z*. In a pristine hybrid gel with a 35% starch content and no stretching history, the 2D FFT pattern was a featureless circular ring (**Fig. 2C**, left panel), indicating that the initial distribution of starch granules was random and isotropic. As the strain was increased, the original uniform distribution evolved into an elliptical pattern with a gap at $k_z$ = 0 and two discontinuous arcs (**Fig. 2C**). When the strain level reached a critical point (*i.e.,* ~100%), another gap appeared at $k_r$ = 0 that further separated the two arcs into four symmetrical islands (the two rightmost arcs are shown in **Fig. 2C**). These hotspots centered at ($k_r^*$, $k_z^*$) with the $k_z^* / k_r^*$ ratio asymptotically approaching a constant value $|k_z^* / k_r^*| \approx 0.8$ (**Fig. 2C**) until the strain reached ~500%. When the initial starch packing fraction was low, the transition at $\varepsilon = 100\%$ was not pronounced during stretching (**fig. S5**). This means that at maximum strain, the average intergranular distance was elongated along the extension direction $z$ (**Figs. 2C-E**). Meanwhile, denser structures of starch granules were formed along the $r$-direction, which was consistent with the overall deformation of the hybrid gel with a positive Poisson's ratio (**fig. S6**). At the beginning of the second stretching (**Figs. 2C**, **2D**) or when the first stretching was over, *i.e.*, at 0% strain, the $k_z^* / k_r^* \approx 1.8$, which represented an 80% increase from the ratio at the initial state of the sample (*i.e.*, 1). Such a sharp transition indicates that strain can trigger granular motion inside the viscoelastic PAA-Alg matrix. After the second stretch, no sharp transition of microstructural reorganization was observed (**fig. S7**) since the granular particles were already pre-aligned after the first cycle (**Fig. 2E**).

To investigate the stretch-induced motion of starch granules, we used X-ray photon correlation spectroscopy (XPCS). A hydrogel sample was stretched three times, keeping the maximum strain constant and we recorded the relaxation dynamics of the starch immediately after the hydrogel reaches the maximum strain (**Fig. 3A**). The starch granules displayed constant motions within the polymeric matrix for at least 120 s after the first stretch (**Fig. 3B**). The motions were less pronounced after the second stretch (**Fig. 3C**) and were negligible after the third stretch (**Fig. 3D**). The relaxation dynamics from the XPCS studies are consistent with the coherent X-ray scattering images (**fig. S8**) and the static X-ray CT results (**Fig. 2**), highlighting that the initial stretch is critical to programming the granular structures within the hydrogel matrix. The arrest of the relaxation dynamics after the second stretch suggests that mechanical programming of the hydrogel composite only requires 1-2 cycles of training.

**Programming of mechanical heterogeneity for robotic skins**

Cellular orientation affects local tissue mechanics. The strain-induced reorganization of starch granules within the hydrogel matrix is reminiscent of directed cell migration over ECM under tissue strain or other external stimuli. We next performed tensile tests to evaluate if changes in granular position can affect the mechanics of our hybrid gel.

In two loading/unloading cycles to the same strain level, the hybrid gel consistently showed significantly reduced hysteresis in the second cycle compared with the first cycle (**Figs. 4A-C** and **Supplementary Fig. 9**). With further repetitive cycles, the hysteresis area stabilized at the low level over the next 49 runs (**Supplementary Fig. 9**). The hysteresis during a loading/unloading cycle reflected an energy dissipation, likely used for the reorientation of the granules. When we increased the maximal strain, the stress-strain curve showed different trends before and after the previous maximal strain point (**Figs. 4B** and **4C**). Specifically, the hysteresis area remained low before the previous maximal strain level, but it become much larger afterwards (**Figs. 4A-C**). This observation suggests that the energy dissipation capability of the starch hybrid gel is dependent on its mechanical training history. Additionally, comparison of the second loading/unloading cycles recorded with different maximum strains (*i.e.*, 200% strain $2^{nd}$ stretch in **Fig. 4A**, 400% strain $2^{nd}$ stretch in **Fig. 4B**, and 600% strain $2^{nd}$ stretch in **Fig. 4C**) highlights that the hydrogel composite becomes generally softer if the maximum strain applied for the first cycle (*i.e.*, 600%) is larger.

This strain history-dependent programming of mechanical properties in hydrogel composites suggests a previously unexplored means of programming mechanical heterogeneity in a single material. Simply, if different levels of strain are applied to different locations across the same starting material, the areas that undergo the largest initial strain will be the most compliant in subsequent operations (**Fig. 4D**). In real tissues such as skin, such heterogeneous mechanical properties allow for different functions and motions in different parts of the body, for example in the joints, the back, and the underside of human hands.

To illustrate how the hydrogel composite could be potentially trained *in situ* for heterogeneous robotic skins, we coated the as-made hydrogel layer over the surfaces of a robotic hand (**Figs. 4D**, **4E**). Bending of the robotic fingers at the joint positions during the first several cycles locally deformed the hydrogels, thus producing softer hydrogel domains around the joints; this compliance enabled smooth and robust finger movement during subsequent cycles (**Fig. 4E** and **Movie 1**).

**Reprogramming and control of microstates**

Local mechanical behavior of biological tissue is usually anisotropic and reprogrammable, characteristics which are associated with adaptable cellular structures and biophysical states. We next explored the possibility of reprogramming and modulating granular microstates inside the hydrogel matrix.

We employed orthogonal stretching and kneading to reprogram the microstates for starch granules (**Fig. 5A**). We investigated three macroscopic mechanical responses: the 1$^{st}$ stretching cycle, which we considered 'WRITE'; further stretching along the same direction after the 1$^{st}$ cycle to access the memory retention; and stretching along an orthogonal direction, which we considered 'OVER-WRITE'.

Due to the substantial starch reorganization, the first stretch and release cycle of a hybrid gel typically dissipated the largest amount of energy (**Fig. 5B** and **figs. S10-13**). Subsequent stretching cycles consistently produced much smaller hysteresis areas that were largely overlapping (**Fig. 5B** and **fig. S13**), suggesting that starch grains were already mechanically trained in a defined state after the first stretching cycle (*i.e.*, 'WRITE' operation). However, after rotating the sample by 90° and stretching it in an orthogonal direction, the hydrogel composite showed a pronounced hysteresis area recovery (**Fig. 5B** and **figs. S10-13**), suggesting mechanical anisotropy. Further stretching in the same 90° direction only yielded small hysteresis areas (**Fig. 5B** and **figs. S10-13**). Likewise, the hysteresis area can be restored by stretching at 180° relative to the initial direction (**Fig. 5B** and **figs. S10-13**). Essentially, we were able to modulate the energy dissipation of the gel simply by controlling the stretching direction, *i.e.,* the gel can be regarded as a mechanical memory device whose energy dissipation level reflects its history (**Fig. 5B**). Finally, randomly applied strains obtained by kneading the sample can rejuvenate the gel (*i.e.*, 'ERASE' operation) and recover its ability to dissipate the maximum energy by resetting the starch organization (**Fig. 5A** and **fig. S14**).

To further demonstrate the controllability of such a mechanical memory device, we studied the dependence of the hysteresis area on the applied strain. We used the hysteresis area during the first cycle at 90° (*i.e., $S_{90°,1st}$*) as the reference area. With increasing strain ($\varepsilon$), the $S_{90°,2nd}/S_{90°,1st}$ ratio decreases (**Fig. 5C**). This suggests that the one-time 'WRITE' is more pronounced at higher $\varepsilon$, consistent with the strain-dependent granular organization (**Fig. 5C** and **figs. S10-13**). Nevertheless, $\varepsilon$ does not significantly affect the $S_{180°, 1st}/S_{90°, 1st}$ ratio (**Fig. 5C**), suggesting that the re-organization of starch granules along an orthogonal direction (*i.e.,* 'OVER-WRITE' operation) can still be effective even though the prior 'WRITE' is enforced by a larger $\varepsilon$.

**Figure 5D** summarizes our results for the three macroscopic mechanical responses encoded by the granular motion and patterning: ***X*** (the 1st stretching cycle) during 'WRITE', ***Y*** (further stretching along the same direction after the 1st cycle) to access the memory retention, and ***Z*** (stretching along an orthogonal direction) during 'OVER-WRITE'. Additionally, kneading 'ERASE's the encoded mesoscale states of starch granules.

To connect the strain-induced evolution of the mesoscale states (**Fig. 2**) to the macroscopic mechanical responses, we performed another set of X-ray micro-tomography on the hybrid gels after orthogonal stretching (**Fig. 5E** and **fig. S15**). The as-made starch-filled hydrogel sample was in an isotropic state ($\alpha$). During the first stretching, an asymmetric state ($\beta$) in the ***k***-space formed and followed the overall shape change of the hybrid gel, which exerted an ellipsoidal 2D Fourier diffractogram that elongated in the ***r***-direction (**Fig. 5E**). The $\gamma$ state was obtained after the unloading process, which was also anisotropic, but the long axis was aligned along the ***z***-direction (**Fig. 5E** and **Fig. 2C**). Interestingly, subsequent X-ray tomography of the starch distribution during stretching in other directions revealed the existence of more intermediate states of $\beta'$ and $\gamma'$, which partially resembled the anisotropic features of $\beta$ and $\gamma$ but with slightly less pronounced asymmetry (**fig. S15**). **Figure 5F** summarizes the general working flow that guides the operation of our mechanically encoded memory device using these basic mesostructural building blocks (**Fig. 5E**, left). Similar to muscle memory formed by motor learning during strength training, the wide-range reconfigurability of the hydrogel composite facilitates its reorganization into multiple forms that can respond to different mechanical impacts.

**Strain-stiffening action**

Strain-stiffening behavior protects biological tissues against external mechanical impact and prevents tissue disintegration. This behavior is largely attributed to the strain-induced dynamic response of the ECM and cellular ensemble. To determine if the observed reorganization of the starch granules (**Fig. 2**) induces a change in the bulk modulus of the composite, we used ultrasound shear wave elastography to directly measure the shear modulus $G$ of the gel under stretching (**Fig. 6, A-C** and **Movies 2, 3**). Typically, the shear modulus of a dense starch suspension increases significantly when the starch granules are jammed(*49, 52*). Given the apparent alignment of the starch structures observed on the X-ray images (**Fig. 2C**), we expected to see a similar increase in shear modulus $G$ during hybrid gel stretching. For the ultrasound experiment, we pushed a transducer against the bottom surface of a stretched gel (**Fig. 6, A** and **B**) and monitored the propagation of focused ultrasound-induced shear waves (**Fig. 6C**). We extrapolated the genuine shear moduli $G$ of individual samples, which follows $G = \rho c_s^2$, where $\rho$ is the gel density and $c_s$ is the shear wave propagation speed. In agreement with the X-ray tomography and bulk tensile test (**Fig. 2C** and **fig. S16**) that showed a critical strain of 100% at the onset of substantial starch aggregation, there was a clear inflection point near $\varepsilon = 100\%$ during the first stretching, after which the gel stiffened nearly fourfold from 34 kPa (at $\varepsilon = 100\%$) to 113 kPa (at $\varepsilon = 200\%$) (**fig. S16**). Furthermore, during the second stretching process, the hybrid gel showed consistently lower values of shear moduli until the maximum strain was reached, although gradual stiffening of the gel remained evident with the increased strain level (**fig. S16**). Collectively, these characterizations led to the conclusion that strain-induced reorganization of the starch granules correlated strongly with the macroscopic mechanical properties of the hydrogel composite.

**Other tissue-like properties**

Besides ***mechanical programmability***, ***anisotropy*** and ***strain-stiffening***, biological tissues have evolved several other remarkable features for sensory or protective functions. For example, the rare combination of softness and toughness in mammalian skin tissue generates an effective barrier to protect internal organs against environment damage. Additionally, when a tissue is wounded, its mechanochemical response and regenerative mechanisms yield timely self-healing. Given the high structural similarity between our starch hybrid gel and biological tissues, we explored other tissues-like properties in the hybrid gel system.

*Toughness*: We first performed a bulk tensile test of the starch hybrid gel to determine its toughness (**Fig. 7A**). While the gel has a tissue-level Young's modulus of ~36 kPa, its stretchability is much higher than that of biological tissues (~7900% for the gel vs ~110% for skin). To dissect how the starch granules and polymeric network composition enable this high toughness, we conducted tests on different control samples (**table S2**). Upon addition of increased concentrations of urea - a known disruptor of hydrogen bonding(*47*) - to the hybrid gel, we observed that the stretchability and toughness of the gel progressively decreased (**Fig. 7A** and **fig. S17**). In the absence of starch, the PAA-Alg hydrogel had much less stretchability than the hybrid gel (**Fig. 7A** and **figs. S17-S18**). Both the fracture strain and the Young's modulus of the hybrid gel increased with higher loadings of starch granules (**fig. S19**). When we replaced the starch with microparticles of silica ($SiO_2$), a structural analog with dimensions similar to those of the starch grains but with chemical and mechanical differences, the stretchability of the gel composite was greatly compromised (**Fig. 7A**). Moreover, when Alg was removed from the matrix, the fracture strain of the hydrogel composite dropped from ~7900% to ~4090% (**Fig. 7A**). Alginate dispersion within the PAA may have lowered the overall crosslinking density of the hydrogel matrix and thus enhanced the granular motion, as evidenced by the decreased Young's modulus (36 kPa *vs*. 75 kPa; **fig. S17**) and enhanced viscoelastic behaviors (**figs. S20** and **S21**) when Alg was incorporated into the hydrogel composite.

*Mechanochemistry*: To further understand the molecular picture behind the toughness of the hybrid gel, we performed attenuated total reflectance Fourier transform infrared (ATR-FTIR) spectroscopy of the gel under strain (**Fig. 7B,** detailed in **fig. S22**). Notably, in addition to the main absorption peak (1637 $cm^{-1}$) of PAA, two shoulder peaks (1606 and 1674 $cm^{-1}$) started to emerge as the strain amplitude increased (**Figs. 7B**). These shoulder peaks were assigned to the sheet-like structures which formed between the side chains of PAA due to chain alignment under stretching. Control experiments on a $SiO_2$-filled gel or pure PAA-Alg did not show any noticeable changes during stretching (**fig. S23**). *In-situ* X-ray scattering confirmed that the intrinsic starch structure was unaltered during stretching (**fig. S24**). These observations suggest that, in the presence of starch granules, interchain hydrogen-bonding network may have been enhanced during the stretching process. Given the applied strain increases the strength of hydrogen bond network through granules-enabled strain 'focusing'(*18*), these results suggest an approach for achieving mechanochemical responses in tissue-like materials.

*Self-healing*: Due to the substantial number of hydrogen bonds induced by the starch granules, we speculated that dynamic bonding may occur rapidly in our hybrid gel even at room temperature, similar to a self-healing tissue(*53-55*). To verify the self-healability of the hybrid gel, we intentionally cut the gel and let it recover under ambient condition. Tensile test results for the hybrid gel showed a gradual recovery of stretchability to ~90% of its original value within 72 hours after the initial cut (**Fig. 7C** and **fig. S25**). Using X-ray CT, no apparent cracks could be observed after the self-healing process (**Fig. 7D**).

*Impact absorption*: Given the large hysteresis observed during the cyclic tensile test (**Fig. 5**), we hypothesized that reorganization of the starch granules would give rise to significant energy dissipation under external impact, reminiscent of human skin. To test our hypothesis, we dropped a steel ball on a suspended film of starch hybrid gel. Although the film underwent an extreme deformation, it maintained its integrity and held the ball steadily (**Fig. 7E**, upper panel and **Movie 4**). In contrast, a film made of PAA-Alg immediately broke when the ball was dropped on it (**Fig. 7E**, lower panel and **Movie 5**). Kymograph analysis of a vertical compression test showed less lateral fluctuation in the hydrogel composite compared with the control PAA-Alg sample (**Fig. 7F-I**, and **Movies 6-8**), further highlighting the energy absorption capability enabled by the granular materials.

## CONCLUSION

We have demonstrated that cellular-scale starch granules can turn conventional hydrogels into tissue-like materials for robotic skins (**Fig. 8**). Mesoscopic granular motions and actions (*e.g.,* mechanochemical stiffening of the hydrogel matrix) can be programmed by global mechanical manipulation. Multiscale interactions between the starch granules and the surrounding hydrogel matrix yielded a number of tissue-like properties (**Fig. 8B**, **table S3**), such as directed motion of granules, mechanical heterogeneity and anisotropy, microstate programmability, strain-stiffening, improved impact absorption and material toughness, mechanochemistry. Importantly, all of these features were dependent on the presence of starch granules *i.e.*, they could not be obtained with PAA-Alg alone. Our work also hints at the physical biology pathways that underlie numerous behaviors in living tissues (*e.g.*, mechanical response of biofilms). Finally, our results present new design considerations for metamaterials or adaptive systems(*56-60*) for soft robotics and bioelectronics(*9, 39, 40, 61, 62*), whose functions can go beyond those of natural systems.

**ACKNOWLEDGMENTS**

**Funding:**

We thank Karen Watters for scientific editing of the manuscript. This work was supported by the US Office of Naval Research (ONR YIP, N000141612530; PECASE, N000141612958) and the National Science Foundation (NSF MRSEC, DMR 1420709). The work made use of the BioCryo facility of Northwestern University's NUANCE Center, which received support from the Soft and Hybrid Nanotechnology Experimental (SHyNE) Resource (NSF ECCS-1542205); the Materials Research Science and Engineering Centers (MRSEC) program (NSF DMR-1720139) at the Materials Research Center; the International Institute for Nanotechnology (IIN); and the State of Illinois, through the IIN. It also made use of the CryoCluster equipment, which received support from the Major Research Instrumentation (MRI) program (NSF DMR-1229693). This research used resources at the Advanced Photon Source, a U.S. Department of Energy (DOE) Office of Science User Facility operated for the DOE Office of Science by Argonne National Laboratory under Contract No. DE-AC02-06CH11357. This research also used resources at the Full-Field X-ray Imaging Beamline (FXI) at 18-ID of the National Synchrotron Light Source, a U.S. Department of Energy (DOE) Office of Science User Facility operated for the DOE Office of Science by Brookhaven National Laboratory under contract No. DE-AC02-98CH10886. We also thank Dr. Tao Sun for providing technical supports.

**Author contributions:**

Y.F. and B.T. conceived the idea. Y.F., E.H., X.-X.Z., Y.J., H.J and B.T. designed the research. Y.F., E.H., X.-X.Z., Y.J., Y.L., J.S., J.W., L.M., X.G., P.G., X.X., H.T., H.Z., X.Z., Q.Z., M.C., Q.Z., Y.G., R. B., Z.M., Z.J., J.Y., carried out the experiments and L.R. performed the numerical simulation. Y.F., E.H., X.-X.Z., Y.J., Y.L. analyzed and interpreted the results. Y.F., E.H., X.-X.Z., Y.J., and B.T drafted the manuscript and all authors contributed to the writing of the manuscript.

**Competing interests:** The authors declare no competing financial interest.

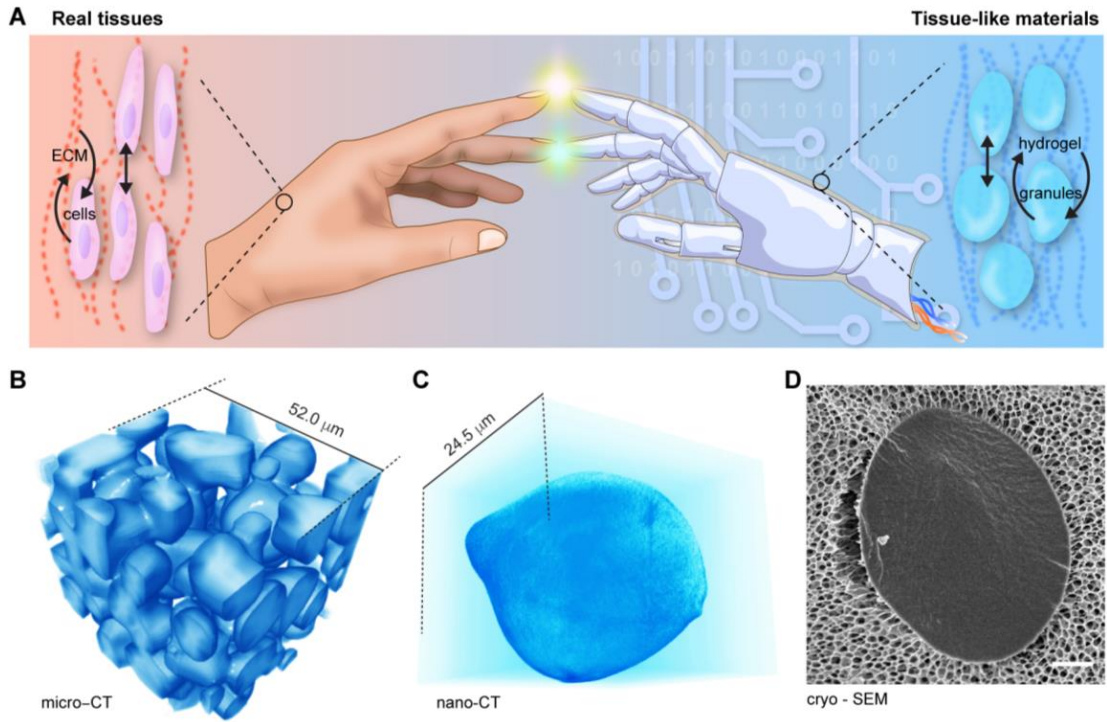

**Fig. 1. Programmable and dynamic granular materials are found to enable tissue-like behaviors in hydrogels.** (**A**) Schematic depicting the active response of biological tissues (or granular hybrid gels) to mechanical stresses through the formation of cellular clusters (or granular aggregates) and preferential alignment of extracellular matrices (or polymeric network). (**B**) and (**C**) Reconstructed X-ray computed tomography (CT) data from a starch hybrid gel showing the irregular shaped single particle (**B**) with nanoscale surface roughness (**C**). (**D**) Cross-sectional scanning electron cryomicroscopy (cryo-SEM) image showing the seamless integration between the starch granule and the surrounding nanostructured polymeric matrix. Scale bar, 1 μm.

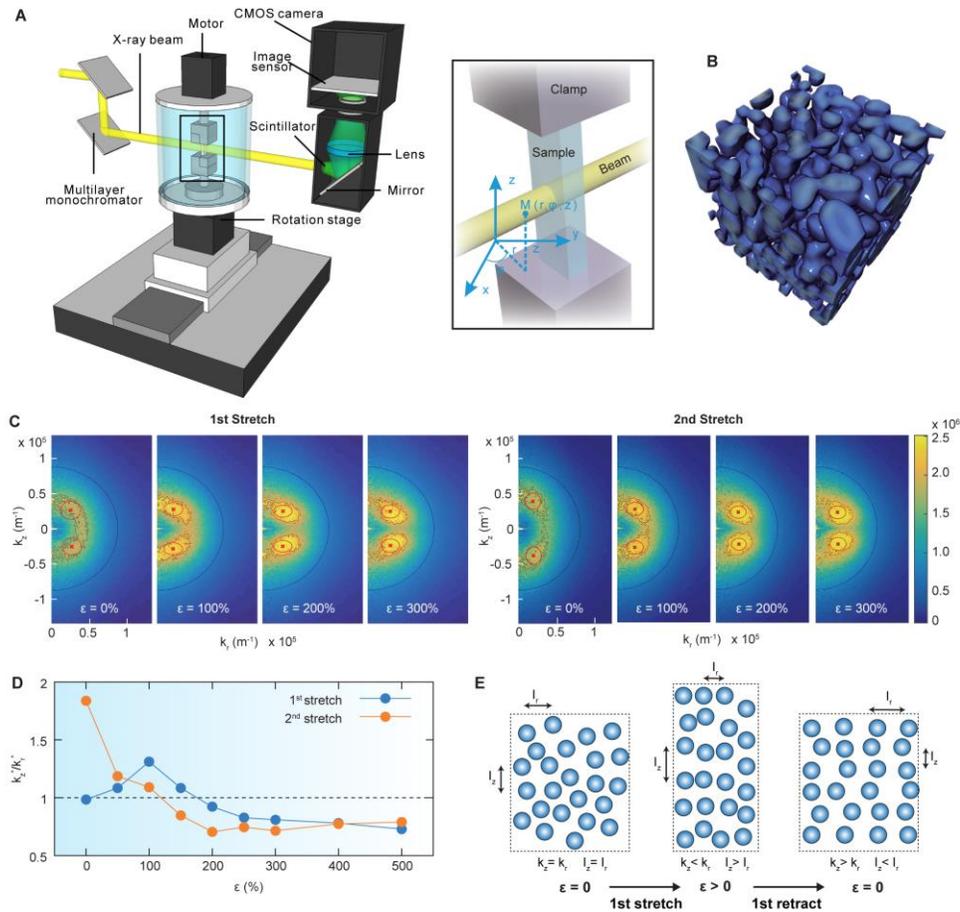

**Fig. 2. Strain can induce mesoscopic motion of starch granules in hydrogel matrix.** (**A**) Schematic of the X-ray micro-CT setup showing the relative orientation of the sample, X-ray beam, and clamps that hold and stretch the hybrid gel sample. (**B**) Three-dimensional (3D) reconstructed image of the spatial distribution of starch granules in the hybrid gel at 500% strain. (**C**) Two-dimensional (2D) fast Fourier transform (FFT) spectra of X-ray images from the same hybrid gel under different strains during the first and second stretching processes. Black circles, spatial frequency corresponding to the average particle diameter of 11 µm. Red circles containing crosses, weighted mean positions of starch granules in the reciprocal space. (**D**) Evolution of starch distribution as a function of strain. The center position changes of the hot spots (red crosses marked in **C**) at different strain levels in the first and second stretching cycles were analyzed and plotted. (**E**) Schematic of starch granules in hybrid gel showing the microscopic alignment under global mechanical manipulation.

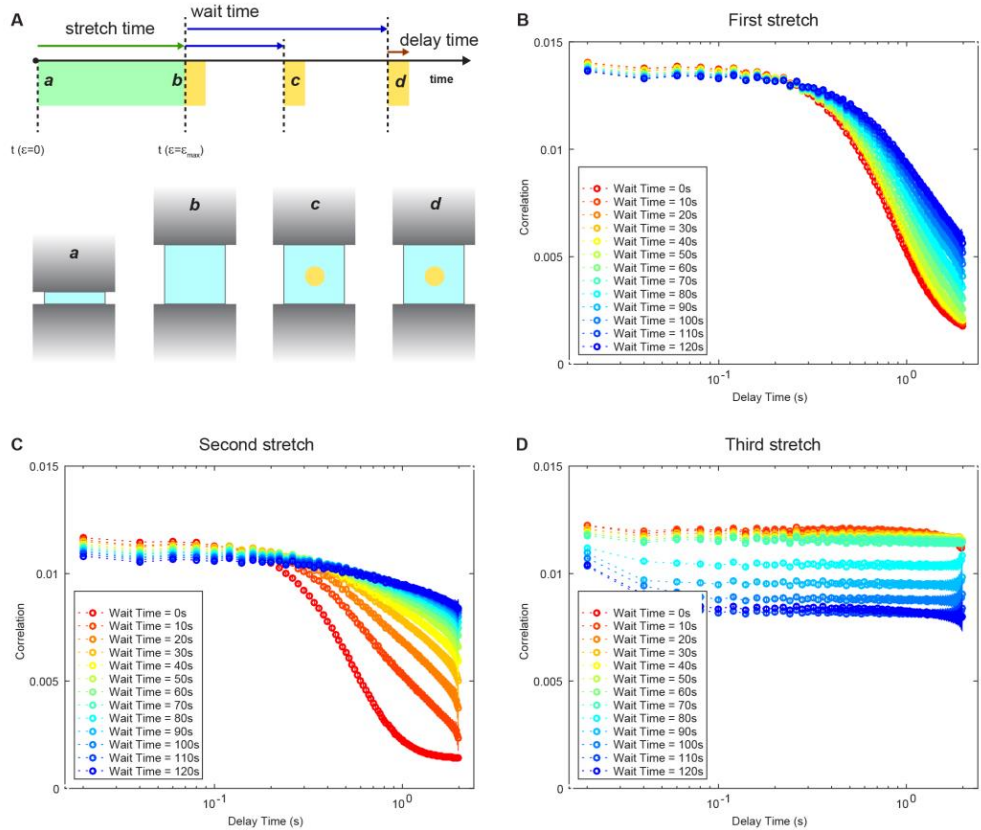

**Fig. 3. X-ray photon correlation spectroscopy characterizes the motion relaxation dynamics of starch granules in hydrogel composite.** (**A**) Schematic diagram illustrating the experimental protocol. A starch hybrid gel sample was stretched three times and the granular relaxation dynamics were probed (yellow areas) immediately after each stretching process using X-ray photon correlation spectroscopy. (**B-D**) The correlation function calculated from near-field, transmission geometry X-ray scattering intensities after the first stretch (**B**), second stretch (**C**), and third stretch (**D**) while the sample is held stationary by the clamps. Each measurement lasted for approximately 120 s. The relaxation of the starch grains immediately after the stretching results in significant rearrangement and dislocation of the particles within a 120 s time window, but then become almost completely arrested within similar time windows with consecutive stretches.

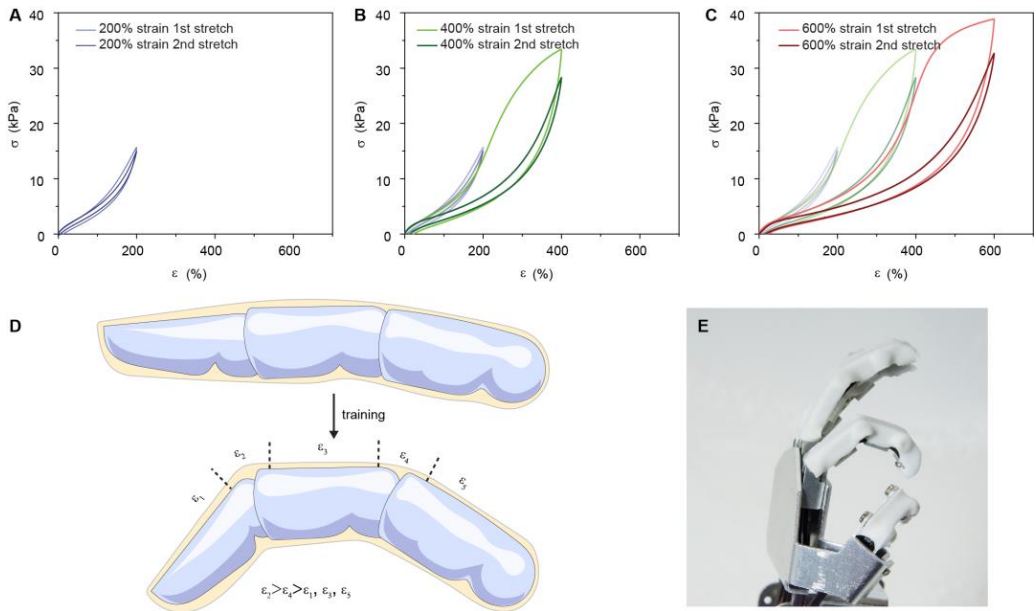

**Fig. 4. Strain-dependent training produces mechanical heterogeneity in the hydrogel composite.** (**A**-**C**), The energy dissipation capability of the starch hybrid gel is dependent on its history of external stress. After the first loading/unloading cycle to a certain strain, the gel consistently showed significantly reduced hysteresis in the second cycle to the same strain level. In addition, the hysteresis area remained low before the previous maximal strain level, but it became much larger afterwards. Maximal strain levels, 200% (**A**); 400% (**B**); 600% (**C**). (**D**) Schematic showing the requirement for heterogeneous mechanical properties in finger skin. (**E**) Starch hybrid gels coated on a robotic hand with heterogeneous compliance after *in situ* mechanical training. These compliant hybrid gels are promising candidates for robotic skin applications.

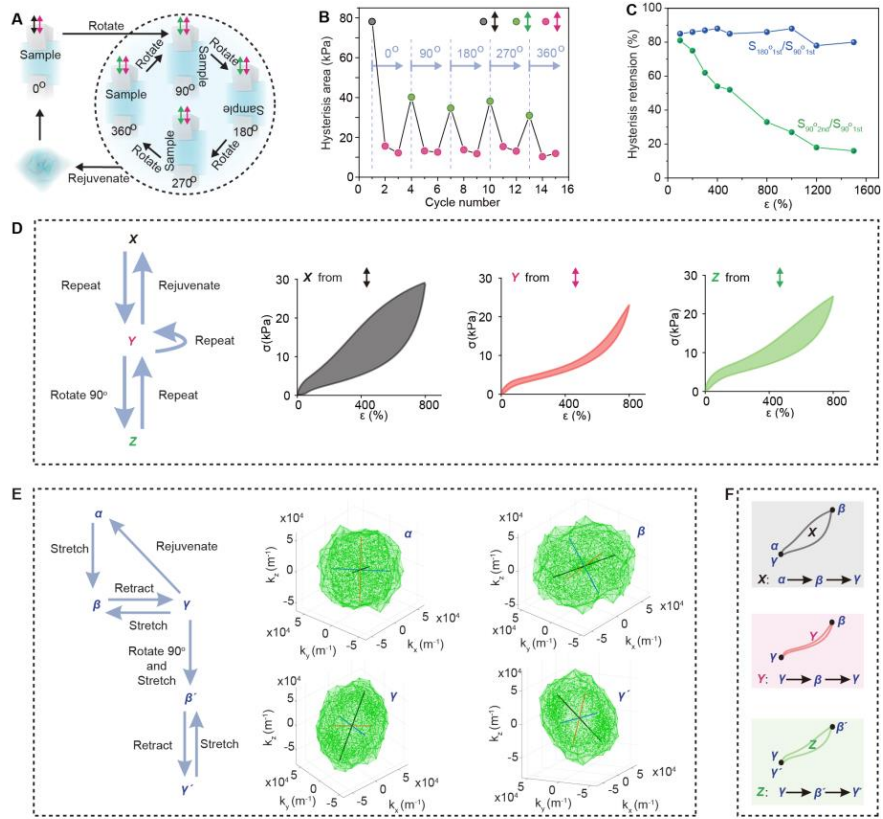

**Fig. 5. Mechanically trained starch granules enable reprogrammability.** (**A**) Schematic of basic reconfiguration steps for the hybrid gel. Two-way arrows represent stretch/relax cycles. Black arrows represent the first cycle on a pristine sample. Pink arrows represent subsequent cycles along the same direction. Green arrows represent the first cycles after each 90° rotation. Blue one-way arrows denote other steps between stretch/relax cycles, such as rotating and kneading. Color coding remains the same throughout **Fig. 5**. (**B**) Energy dissipation, as manifested by the hysteresis area, in a starch hybrid gel changes according to rotation of stretching directions. Sample was stretched to $\varepsilon = 800\%$ in each cycle. (**C**) Quantitative analysis of the gel memory effect with respect to strain level. Data from the first and second cycles following 90° rotation ($S_{90°, 1st}$, [Reference] and $S_{90°, 2nd}$, [Low state], respectively), and the first cycling following 180° rotation (*i.e.*, $S_{180°, 1st}$, High state) were analyzed. (**D**) Macroscopic responses of the mechanical memory device and corresponding stress-strain curves. (**E**) Microscopic states of the starch granules based on 3D FFT patterns of particle configurations from the X-ray micro-CT data. (**F**) Macroscopic responses and microscopic structures are highly correlated.

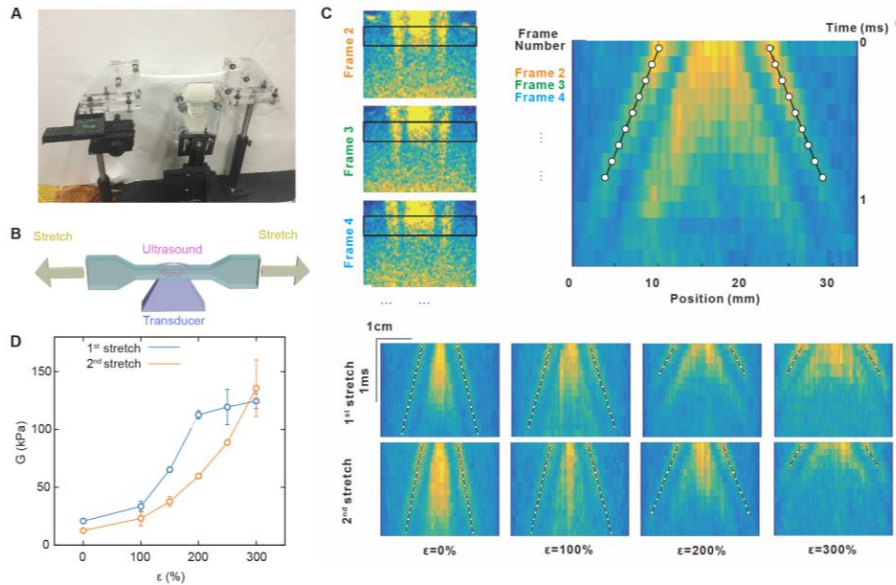

**Fig. 6. Characterization of the hydrogel composite under mechanical manipulation with ultrasound shear wave elastography.** (**A**) and (**B**) Photograph and schematic of ultrasound shear wave elastography system. An ultrasound transducer was attached to the bottom of a starch hybrid gel. The starch granules serve as the signal tracer and enhancer for ultrasound imaging. (**C**) Propagation of shear waves on both sides of the focal point. Images were obtained by calculating the differences between in-phase and quadrature data from adjacent frames. Bright color represents regions with non-zero displacement between adjacent frames. Dashed black boxes denote areas used to generate averaged motion images for each frame. Averaged motion images from each frame were stacked together to highlight the shear wave propagation. Positions of shear wave fronts - one on each side of the middle focal point - are labeled with open circles. Shear wave positions as a function of strain and time for the hybrid gel during the first and second stretching processes. Each panel is an average over three repeated measurements. (**D**) Shear moduli calculated from shear wave speeds at different strains for two stretching cycles. Three measurements were performed at each strain for statistical analysis. Open circles represent mean values of the shear modulus. Error bars denote standard deviations.

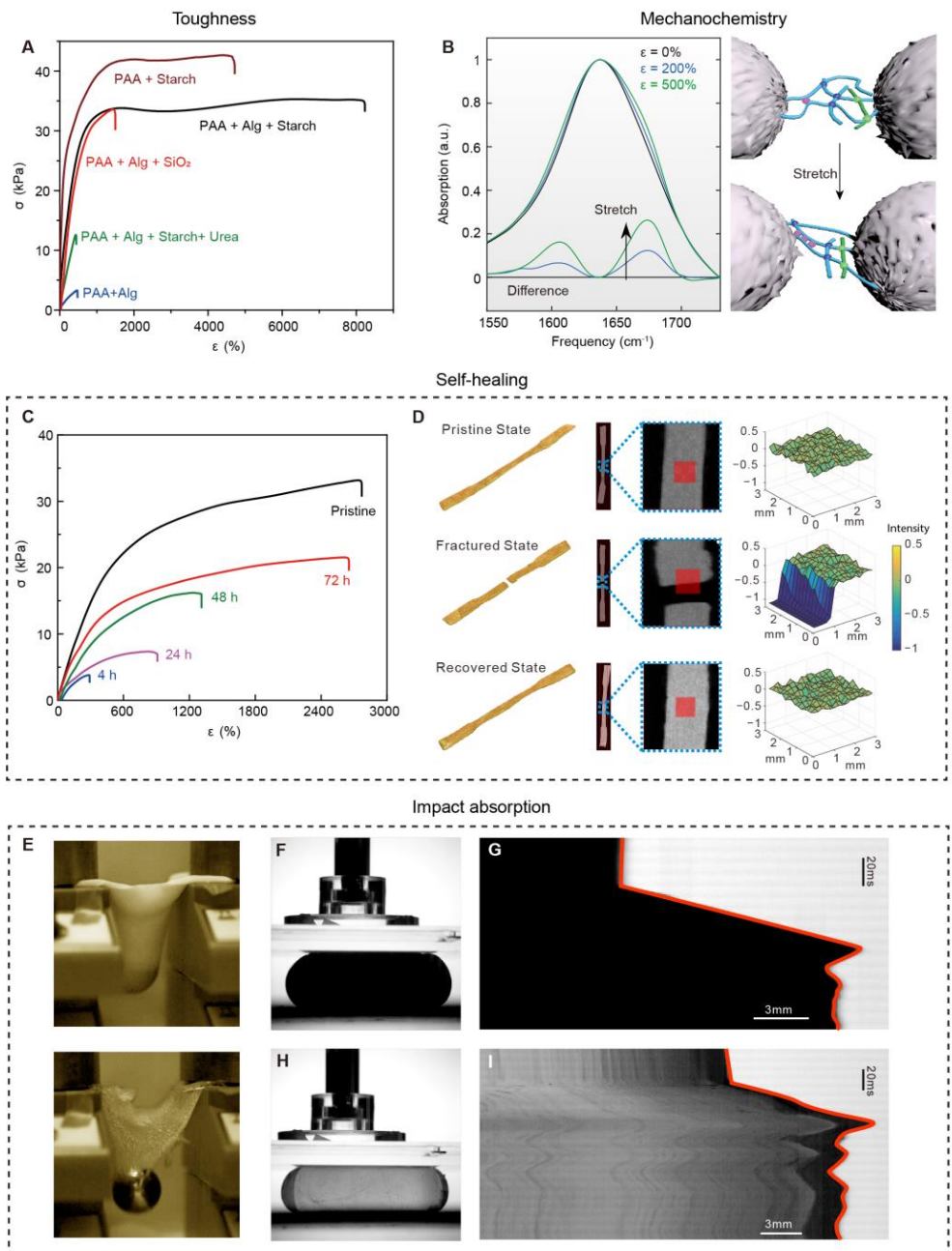

**Fig. 7. Starch granules yield many other tissue-like properties in the hydrogel composite.** (**A**) **Toughness**. Stress-strain curves illustrating the roles of individual components in determining material stretchability. Nominal engineering stress (σ) was defined as the loading force divided by the cross-section area. Extension rates for all samples were 75 mm/min. (**B**) **Mechanochemistry**. Attenuated total reflectance Fourier transform infrared spectra of the same starch-filled gel at different strain levels. Difference spectra (scale factor, 3) are plotted at the bottom with respect to the unstrained sample (black trace). Blue and cyan traces represent data from ε = 200% and ε =

500%, respectively. Schematic depicting enhancement of interchain hydrogen bonds (magenta dots) between adjacent PAA side-chain groups upon stretching. (**C**) **Self-healing**. Stress-strain curves demonstrating the self-healing behavior of the starch hybrid gel. Initial malleability and Young's modulus were gradually recovered after extended healing under ambient conditions. (**D**) **Self-healing**. X-ray tomography data (3D volume rendering, 2D view and surface profiles) from pristine, fractured, and recovered states of a hybrid gel (27 wt% starch in PAA+Alg). Red squares mark regions used for profile analysis. (**E-I**) Ball drop (**E**) and vertical impact (**F** and **H**) tests showing the significantly enhanced impact absorption of the hydrogel composite versus the control PAA+Alg gel. Kymographs in **G** and **I** show the lateral fluctuations.

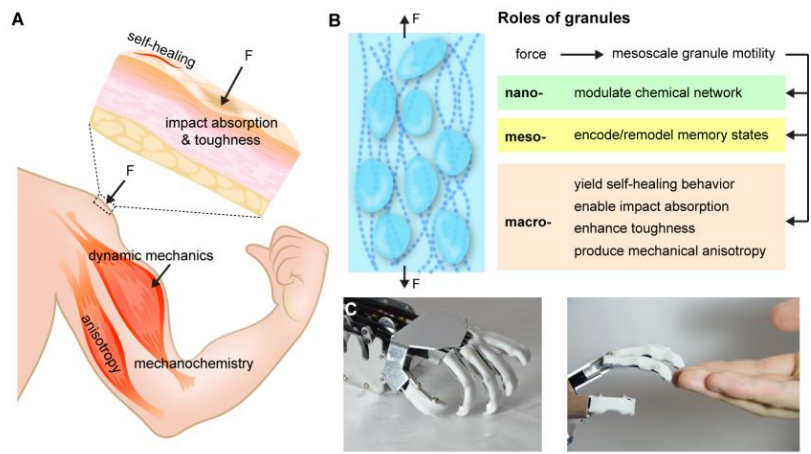

**Figure. 8. Summary of key features enabled by starch granules for tissue-like materials.** (**A**) Biological tissues exhibit an array of unique features including self healability and impact absorption from skin tissue, and dynamic mechanics, anisotropy, and trainability and memory from skeletal muscle. (**B**) Table summarizing the key features enabled by the granular materials across multiple length scales for tissue-like materials. (**C**) Photographs of robotic hands partially coated with starch hybrid gel as artificial skin.

# Supplementary Information

**1. Materials and Methods**

**1.1 Materials**

The acrylamide (AA) monomer, sodium alginate (Alg), crosslinker methylenebisacrylamide (MBAA), unmodified starch from wheat, initiator ammonium persulfate (APS) and accelerator N,N,N',N'-tetramethylenediamine (TEMED) were all purchased from Sigma Aldrich. The control group diols functionalized silica was obtained from Silicycle.

Wheat starch granules have two distinct classes A-type and B-type, which differ in size and shape. The A-type granules are large, lenticular particles, and the B-types are small, spherical particles. The volume ratio of A-type and B-type is approximately 73:27. Their mean diameters are 14.1 ± 0.6 μm and 4.12 ± 0.28 μm, respectively. So, the mean diameter by volume is about 11.4 μm (Ref. 1-3).

**1.2 Synthesis of starch and hydrogel hybrid gels.**

The starch hybrid gel networks were synthesized by mixing the composites following the stoichiometries in **Table S1**, and we also synthesized the several hybrid gels as control groups in **Table S2**.

To prepare the highly stretchable 43 wt% starch hybrid gel networks, we chose Category 1 in **Table S1**. First, we added the wheat starch powder (200 g) in sterilized water (250 g), and the suspension was stirred for 12 hours on a hotplate with stabilized temperature and humidity controlled at 25 °C to ensure the hydration of the starch particles. Then, Alg, AA, APS and MBAA were sequentially added to the suspension and stirred for an additional 48 hours. Unless otherwise indicated, the weight percentage of each component used in the hybrid gel were kept as Alg (0.2 wt%), AA (3.42 wt%), APS (0.11 wt%) and MBAA (0.01 wt%). To accelerate the polymerization, the accelerator TEMED (0.02 wt%) was added into the well-dispersed suspension and stirred for 2-5 min. The suspension was transferred into acrylic molds and sealed with acrylic cover slides for gelation. The hybrid gel was fully polymerized after 3-5 hours.

### 1.3 Mechanical test for the multiple composite gel.

We prepared five types of materials for the tensile test, including polyacrylamide (PAA), PAA + Alg, SiO$_2$+PAA+Alg, and wheat starch particles (16 wt%, 20 wt%, 27 wt%, 35 wt%, 40 wt%, 43 wt%) + PAA + Alg as indicated in **figs. S17, 19**. Dogbone-shaped samples were prepared using casting molds. We followed the ASTM standard (50 × 10 × 6 mm$^3$) to prepare the molds. All of the as-prepared samples were taken out of the molds with care to prevent stretching because the training at large strain is not recoverable by thermal motion (**fig. S16**). After being taken out, the samples were wrapped with paraffin film until the moment before the mechanical testing to keep moist.

We ran tensile tests on two material testers ZwickRoell zwickiLine Z0.5 (100 N loading cell) and Instron 5800 (500 N loading cell) at room temperature. In each test, the two ends of the dogbone-shaped sample were clamped to the material tester via softened interfaces (with rubber tape) to avoid slippage, and the stretching was along the vertical direction. In each test cycle, the sample was stretched from the initial length $L_i$ to the final length $L_f$ at speed $U$ and returned to the original length $L_i$ at the same speed. Here we calculated engineering strain $\varepsilon$, strain rate $\dot{\varepsilon}$ and engineering stress $\sigma$ using

$$\varepsilon = (L_f - L_i)/L_i, \tag{S1}$$

$$\dot{\varepsilon} = U/L_i,$$

$$\sigma = F/A_0,$$

where $F$ is the tensile force, and $A_0$ is the initial cross-sectional area of the sample. The speed $U$ that we used in most experiments was between 60 mm/min and 120 mm/min, and a typical speed was 75mm/min. Multiple measurements were performed for each set of parameters with different samples.

The tensile creep test was conducted using the creep model by ZwickRoell zwickiLine Z0.5 (100 N loading cell). The bone-shaped polymer films ($50 \times 10 \times 6$ mm$^3$) were loaded at constant stresses ranging from 3 kPa to 50 kPa with the loading rate of 75 mm/min.

In the fatigue test, the dogbone-shaped sample ($50 \times 10 \times 6$ mm$^3$) was cyclically stretched and released by ZwickRoell zwickiLine Z0.5 (100 N loading cell) with a maximum strain of $\varepsilon = 800\%$ for 100 cycles with a constant loading rate 75mm/min.

We investigated the recovery efficiency ($\eta$) via the unidirectional tensile test using ZwickRoell zwickiLine Z0.5 (100 N loading cell) Zwick. For the sample preparation, we used the 27 wt% starch hybrid gel as our testing sample ($50 \times 6 \times 4.5$ mm$^3$). We cut the dogbone sample into two pieces and compressed the two separated pieces together. Then the pretreated sample was removed from dogbone mold and sealed by a paraffin film. Later, we tested the healed sample at different durations with the tensile test, such as 4 hours, 24 hours, 48 hours, and 72 hours. All of the experiments were performed under constant humidity at room temperature (25 °C). To evaluate the recovery, we compared the fractured strain and modulus between the pristine sample and the recovered sample. We found that the recovered sample at 72 hours performs ~ 90% recovery of fractured strain and ~ 80% recovery of modulus. We measured four samples for each duration. All the strain-stress measurements were achieved under a strain rate 75mm/min at room temperature.

To verify the reconfigurable energy dissipation, we prepared a square-shaped sample of 35 wt% starch hybrid gel. The dimensions ($x$, $y$, $z$) of the sample were 20 cm, 4 cm, and 20 cm, respectively. We clamped the samples with silicon grids along the $z$-direction at the edges. The sample was prepared in such a way that the grid occupied 1/3 of the spacing while sitting along the center axis. Then, we cyclically loaded the sample three times in the $z$-direction. Then, we removed the trained sample from the clamps, rotated the sample 90° around the $y$-axis, and clamped it back along $x$-direction (labeled as "90°" in **fig. S15**). We performed tests on the rotated sample following the previous procedure. The same tests were also performed with samples that were rotated twice (stretched in $z - x - z$ directions, labelled as "180°") and three times (stretched in $z - x - z - x$ directions, labelled as "270°"). In the angular loading experiment, we also test the angular performance at different strain($\varepsilon$), such as 100%, 200%, 300%, 400%, 500%, 800%, 1000%, 1200%, and 1500%. All the strain-stress measurements were achieved at stretching speed $U$ = 75mm/min at room temperature [4,5].

**1.4 TGA measurements for the hybrid gel.**

Thermogravimetric analysis (TGA) measurements were conducted using a TA Instruments Discovery TGA equipped with an infrared furnace, auto-sampler, and gas delivery module. Nitrogen was used as the purge gas at a flow rate of 25 mL/min. Temperature ramp measurements were performed at 20 °C/min from ambient to 700 °C using standard platinum crucibles.

**1.5 Rheometry.**

Rheological studies were conducted using a TA Instruments ARES-G2 separate motor-transducer rotational rheometer equipped with an Advanced Peltier System (APS) for temperature control. All measurements were performed using an APS flat plate bottom geometry and a 25 mm diameter parallel plate upper geometry with a nominal gap = 1 mm. A solvent trap was used to prevent water evaporation and sample drying during measurement. Samples of diameter = 25 mm were mounted onto the APS flat plate, and the upper geometry was brought into contact under a small applied compressive force of 0.2 N. Frequency sweep measurements were performed over the range 0.1–100 Hz under strain amplitudes of 5% to characterize the frequency-dependent complex shear modulus $G^*(\omega)$ of each sample. All measurements were carried out in the linear viscoelastic region, as was determined by strain amplitude sweeps conducted at frequency = 1 Hz.

**1.6 Cryogenic scanning electron microscopy (cryo-SEM).**

The starch, PAA+Alg gel, and 35 wt% starch hybrid gel samples were dissected into small pieces with a razor blade and placed in the recess of a sample carrier (Type "B", #242-200, Technotrade) for the high-pressure freezer (HPM100, Leica). 20% Dextran (#31389, Sigma) in $H_2O$ was used as a filler, and the flat side of another sample carrier was placed on top of the sample before the assembly was high-pressure frozen. Samples were stored in liquid nitrogen until further processing for cryo-SEM. In other cases, dissected sample pieces were mounted on aluminum sample stubs and frozen by plunging into liquid ethane before storage in liquid nitrogen.

For cryo-SEM, the frozen samples carriers or the sample stubs were transferred into a sample loading dock and mounted under liquid nitrogen onto a cryo-SEM sample holder. A precooled cryo shuttle (VCT100, Leica) was used to transfer the cryo-SEM sample holder with the mounted samples into a cryo etching/coating system (ACE600, Leica). Samples were freeze-fractured with a cold blade in the ACE600. For etching, the temperature of the cryo-SEM sample holder in the ACE600 was raised to -95 °C for 6 minutes in a vacuum of $2 \times 10^{-6}$ mbar. After etching, the temperature was decreased to -120 °C and the samples were coated with 5 nm platinum followed by 10 nm of carbon. The cryo-SEM sample holder with the coated samples was loaded into the precooled cryo shuttle and inserted into the SEM (S-4800-II, Hitachi) with a precooled cryo stage

(Leica). The samples were imaged at a temperature of -110 °C with an acceleration voltage of 2 kV.

### 1.7 Micro-computed tomography (micro-CT).

Micro-CT images of 27% starch hybrid gel samples were performed on the XCUBE (Molecubes NV, Belgium) by the Integrated Small Animal Imaging Research Resource (iSAIRR) at The University of Chicago. Spiral high-resolution CT acquisitions were performed with an X-ray source of 50 kVp and 200 µA. Volumetric CT images were reconstructed in a 350 × 350 × 840 format with voxel dimensions of 200 µm$^3$. Images were analyzed using AMIRA 6.4 (Thermo Fisher Scientific, USA), VivoQuant 3.5 patch 2 (InviCRO, LLC, USA), and MATLAB 2015a (MathWorks, USA).

### 1.8 Infrared Spectroscopy.

The sample temperature was monitored with a Phidgets K-type thermocouple attached to the brass jacket. ATR-FTIR spectra were measured using Bruker Platinum ATR.. $H_2O$ was used in **Fig. 7B**, **figs. S22** and **S23**.

### 1.9 Extended Wide-angle X-ray Scattering (WAXS).

Synchrotron WAXS measurements with extended $q$ range (up to 2.0 Å$^{-1}$) were performed at the sector 12-ID-D of the Advanced Photon Source at Argonne National Laboratory. The pristine unstretched and stretched samples (35 wt% starch hybrid gel) were held on a customized mechanical force mounting stage and transmitted by an incident X-ray beam with a 0.6199 Å wavelength and a 0.2 × 2 mm beam profile. The WAXS data was integrated by the 2D scattering patterns that were collected by a Dectris Pilatus 100K photon-counting area detector placed downstream at 394 mm away from the samples. The extended range WAXS data complement the SAXS data of the same type of samples.

### 1.10 Small- and wide-angle X-ray scattering (SAXS/WAXS) experiments.

Simultaneous small- and wide-angle X-ray scattering (SAXS/WAXS) measurements were conducted at Beamline 12-ID-B of the Advanced Photon Source (APS), Argonne National Laboratory (ANL). The wavelength of the X-ray beam was 0.8856 Å, and the beam size was 0.20(H) × 0.10(V) mm$^2$. Two Pilatus X-ray detectors (Dectris Ltd, Switzerland), Pilatus 2M for SAXS and 300K for WAXS, were used for simultaneous SAXS/WAXS measurements. The sample-to-detector distances were set to cover the scattering momentum transfer, $q$, from 0.005 to 2.7 Å$^{-1}$ without gap. The exposure time was set in the range of 0.5-1.0 s to achieve a good signal-to-noise ratio but without detectable radiation damages. The $q$ value calibration was performed using silver behenate prior to measurements. The isotropic 2-D images were converted to 1-D scattering profiles using the Matlab software package developed at the beamline. A Linkam THMS600 heating/cooling stage (Linkam Scientific, UK) was used for the *in situ* heating experiments. The temperature accuracy is within ±0.1 ºC. The SAXS/WAXS data were taken 5 minutes after the heater reaches the set temperature. Water swelling experiments were carried out as the following: 35 wt% starch hybrid gel (dry) was put in a quartz capillary of 2 mm diameter; the X-ray beam was put about 1mm below the top position of the sample, in such way that water added from the top would have good contact with the sample; water was added using a syringe pump which was controlled remotely. SAXS/WAXS data collection was started immediately after adding water to the starch sample.

### 1.11 Shear wave elastography with ultrasound.

In a solid material, the propagation speed of shear wave $c_S$ is a function of the shear modulus $G$ and density $\rho$ of the solid,

$$c_S = \sqrt{G/\rho}. \qquad \text{S2}$$

For soft gels like our samples, a shear wave can be generated and imaged by a medical ultrasound system. Here we used a Verasonics Vantage 128 research ultrasound platform to characterize the mechanical properties of the hybrid gels via shear wave elastography. As schematically illustrated in **Fig. 6B**, the ultrasound transducer was pushed against the tested material, and they were coupled by a layer of ultrasound gel. First, the ultrasound beam was at high power and focused to push the sample with an acoustic radiation force

$$F_{AR} = \frac{2\alpha I}{c}, \qquad \text{S3}$$

where $\alpha$ is the attenuation factor in the material, $I$ is the intensity the sound beam, and $c$ is the speed of sound. Then the high-power beam was turned off, and the ultrasound was immediately switched to the low-power imaging mode and took 500 consecutive images at 10,000 frames per second. To detect small vibrations in the sample, we calculated the difference between adjacent frames

$$\Delta I = |I_{n+1} - I_n|, \qquad \text{S4}$$

where $I_n$ is the in-phase and quadrature data of the nth frame. With $\Delta I$, we can directly visualize propagation of shear waves and track their motion, and thus we obtain the shear modulus of the sample via Eqn. S2. This experimental method was calibrated with gelatin blocks[6].

**1.12 X-ray tomography.**

The micro-tomography experiments were done at 2-BM beamline of Advanced Photon Source at Argonne National Laboratory. The X-ray energy was set to 22 keV. The imaging device was a PCO.edge 5.5 s CMOS camera coupled with a Mitutoyo long working distance 5 × lens and 20 µm thick LuAG:Ce scintillator. A mechanical loading device with two grips were used to stretch dogbone-shaped dough samples in *in situ* micro-tomography experiments. An actuator was used to adjust the gap between the two grips. The maximum travel range of the actuator is 35 mm. The tomography measurements of a sample started with the sample at rest state. The sample was then stretched step by step and scanned after each stretching step. Before a tomography scan started, the sample was relaxed for 15 minutes in order to reduce motions of the starch particles during the tomography scan. The loading device was completely sealed, and a piece of wet foam was inside to keep the sample in a 100% humidity environment. Tomography data was reconstructed with Tomopy. Single distance phase retrieval and ring removal were performed on the raw data before tomography reconstruction. The reconstructions were done with gridrec algorithms. The reconstructed tomography data has isotropic 1.3 µm in all three dimensions. The measured volume in a single tomography scan was a cylindrical volume of 1.6 mm diameter and 1.4 mm height.

To extract information on particle configuration and orientation, we performed a fast Fourier transform (FFT) on the reconstructed images. First, the 2D grey-scale images are combined as a 3D array; then we applied a 3D FFT to the image array. Since the system has rotational symmetry along the extension direction z, we average the 3D Fourier spectrum $(k_x, k_y, k_z)$ in the azimuthal direction and obtain its 2D projection $(k_r, k_z)$. To present the variation of particle configuration directly, we choose the 500 points with the highest intensity on the 2D diagram and compute their center $(k_r^*, k_z^*)$ using intensity as weight. If $k_z^*/k_r^* < 1$, the average length scale of the particle configuration is larger in the extension direction compared to the radial direction, and vice versa.

The nano-tomography measurements were conducted with the transmission X-ray microscope[7] at FXI (18-ID) beamline of National Synchrotron Light Source II at Brookhaven National Laboratory. The scan was done at 6 keV X-ray energy and a 30 nm outmost zone width zone plate. The starch particle samples were mounted on a thin Kapton tube and rotated by an air-bearing rotation stage. X-ray projection images were acquired at 14xy angles in 180° range. The tomography reconstruction was done with Tomopy. Single distance phase retrieval and ring removal were performed on the raw data before tomography reconstruction. The reconstructions were done with gridrec algorithms.

**1.13 Particle rearrangement simulation under tension.**

We developed a simple model to capture how the particle configuration evolves under uniaxial extension and tested it with numerical simulations. The model is this: At small deformation, the particles move affinely with the elastic polymer matrix. However, when two particles touch, they repel each other, thus generating non-affine rearrangement in the system. Guided by this idea, we simulate the particle rearrangement when the material is stretched along the z-direction with LAMMPS. The system contains 100,000 spherical particles. The particles are monodispersed, with a diameter ($d$) of 1.0 cm, density $10^3$ g/cm$^3$, and bulk modulus ~ $1\times10^9$ Pa. They are placed in a square, periodic simulation box of side length 56 cm, so that the initial packing fraction is ~ 0.3. The particles interact via a Hertzian potential with no dissipation or friction.

The procedure of the simulation is:

1. Prepare randomly distributed spheres in a three-dimensional cubic box at packing fraction $\phi$.
2. Deform the box with small incremental strain $\varepsilon$ in the z-direction. The center of each sphere moves affinely with the bulk material as

$$\begin{pmatrix} x' \\ y' \\ z' \end{pmatrix} = \begin{pmatrix} 1 - \varepsilon/2 & 0 & 0 \\ 0 & 1 - \varepsilon/2 & 0 \\ 0 & 0 & 1 + \varepsilon \end{pmatrix} \begin{pmatrix} x \\ y \\ z \end{pmatrix}.$$

3. After every step of deformation, detect if there are any overlapping any particles. If there is no particle overlapping, repeat step 2. Otherwise, let the contacting particles repel each other and relax until there is no longer any overlap.

4. Keep stretching until the system reaches a pre-set final strain. Record the positions of each particle.

After the desired strain has been reached, the process is reversed, using an identical procedure. The total strain of the system changed from $\varepsilon = 0\%$ at the beginning to $\varepsilon = 500\%$, then back to $\varepsilon = 0\%$. We stretched the system one more time in the same direction to strain $\varepsilon = 500\%$, then returned it to $\varepsilon = 0\%$.

In the preparation stage of the simulation, the spheres are distributed randomly in the simulation box and allowed to relax. This is achieved by removing all kinetic energy from the system everyone millisecond until no spheres are contacting, where the time interval is chosen to be sufficiently small so that the spheres are moving quasi-statically. The detailed steps are as follows:

1) Generate random configuration of $N$ (100,000) spheres of a certain stiffness, radius, and density

2) Quench:

   a) Wait time $t_1$.

   b) Remove all kinetic energy.

   c) Wait time $t_2$ (usually more than $t_1$).

   d) Remove all kinetic energy.

   e) Loop back to step a).

f) Break loop when total kinetic energy in system after $t_2$ becomes zero.

In the deformation stage, the following protocol is used:

1) The box is lengthened by 0.1 mm (*i.e.,* 1% of the sphere diameter) in the *z*-dimension and shrunk in the *x* and *y* dimensions to preserve the total volume. The sphere positions are remapped from the previous box dimensions.

2) The sphere positions and stresses are recorded.

3) The system is allowed to relax. The kinetic energy of the system is removed every 0.01 to 0.25 milliseconds, where the time interval is chosen to ensure that the spheres move quasi-statically with respect to each other.

4) The relaxation is allowed to continue until the number of contacting spheres is less than 1.5% of the total and the total average stress per sphere is less than 15 kPa.

5) The final sphere positions and stresses are recorded.

**1.16 Analysis of the simulation results.**

The MD simulations we ran provided positions of every particle in space. To reveal the underlying structure of the particle configuration and be able to compare with the experimental data directly, we analyzed the simulation results in the same way as the X-ray data. After the simulation, we transferred the particle position information to a 3D binary matrix, where 1 represents the region inside the spheres, and 0 represents the outside. We then applied the same data analysis process used for the X-ray data to analyze these simulated results. Since the actual length scale of the particles does not affect the result, we set it to be 1.

**1.17 Analysis of the X-ray tomography data in 3D k-space.**

The X-ray tomography experiments provided 3D reconstructed image files. To obtain the information in k-space, firstly we crop a 400×400×400 (pixel) volume from the original images. The length scale of each pixel was 1.3 µm, so the maximum spatial frequency we obtained in k-space was $3.85 \times 10^5$ m$^{-1}$. To make things easier, we simply used the inverse of length scale as $k$ in this paper, without multiplying $2\pi$.

To present the FFT data in 3D $k$-space, we plot out the 5,000 points with the highest magnitudes. The outer envelope surface of these points is shown in **fig. S15**. The shape of this point set can be approximated as an ellipsoid centered at (0, 0, 0). The fitted ellipsoid can be described by the function

$$\boldsymbol{x}^T K \boldsymbol{x} = 1, \qquad \qquad \text{S5}$$

where $\boldsymbol{x}$ is a vector that points to the surface of the ellipsoid and K is a positive definite matrix. The lengths and orientations of the three principal axes are obtained from the eigenvalues and eigenvectors of $K$, respectively. Their lengths $k_1$, $k_2$, and $k_3$ ($k_1 > k_2 > k_3$) at different strains are shown in **fig. S15** for the first and second stretching processes of the sample, respectively. The trend is identical to what we observed in the averaged 2D Fourier spectra. In the first stretching process, the system started from a more isotropic state, and evolved towards an anisotropic final state. When the strain returned to zero, the system did not recover to the isotropic initial state, but instead, developed into a vertical ellipsoid. When the sample was stretched again, the overall anisotropy went down instead of up when the strain was small. We look at the orientation of this ellipsoid by defining the angle between its longest principal axis and the positive direction of the z-axis, defined as $\theta_L$. As shown in **fig. S15**, for strain $\varepsilon > 50\%$, the angle $\theta_L$ was always around 90°. This means that it was always elongated perpendicular to the stretching direction. However, at the beginning of the second stretching process, the angle was very small, meaning that then the ellipsoid was elongated in the *z*-direction. Both agree with the 2D Fourier spectra discussed in the main text. For the state at $\varepsilon = 0$ in the first stretching process, since the points formed an approximately spherical shape and the three axes had similar lengths, $\theta_L$ was rather random.

**fig. S15** was from samples stretched in the *z*-direction to different strains. These data were used as a benchmark to test this analysis method. We then applied it to analyze the data obtained from samples stretched in perpendicular directions. The experimental procedure was this: we prepared four hybrid gel samples S1, S2, S3, and S4. S1 was stretched in the *z*-directions only. S2 was stretched in the *z*-direction, and then in *y*-direction. S3 was stretched in *z*, in *y*, and then in *z*. S4 was stretched in *z*, in *y*, in *z* again, and in *y* again. After the stretching processes, all the samples returned to 0% strain and were imaged with X-ray. The states that we looked at were $\gamma$ and $\gamma'$ states corresponding to **Fig. 5** in the main text. The resulting shapes of the ellipsoids are shown in **fig. S15**. The angles label the direction of the last stretching (S1 – 0°, S2 – 90°, S3 – 180°, S4 – 270°). We can see that $k_1$ decreased a little, while $k_2$ and $k_3$ showed zigzag shapes. This means that shape-wise, the state $\gamma'$ is a slightly different state as compared to $\gamma$, and approximately, after two consecutive rotations of 90°, the system could return to a state that is closer to $\gamma$. For S1, S2, S3, and S4, the corresponding end states were $\gamma \rightarrow \gamma' \rightarrow \gamma \rightarrow \gamma'$.

Lastly, we looked at the rejuvenation process labeled in **Fig. 5** of the main text. We found that by far the most effective method is to knead the sample (by hand), according to the stress-strain curves shown in **fig. S15**. In **fig. S14**, we tried to look at the end states $\gamma$ before and after kneading. We expected zigzag curves that jump between the anisotropic $\gamma$ state and a more isotropic state, which is closer to the $\alpha$ state shown in **Fig. 5**. However, we saw a trend that was a more monotonic transition. We suspect that this is due to the limitation of kneading by hand, which was hard to control. However, the general trend agrees with what we expected: the ellipsoid in *k*-space did turn into a more isotropic shape[8].

## 2. Supplementary Figures

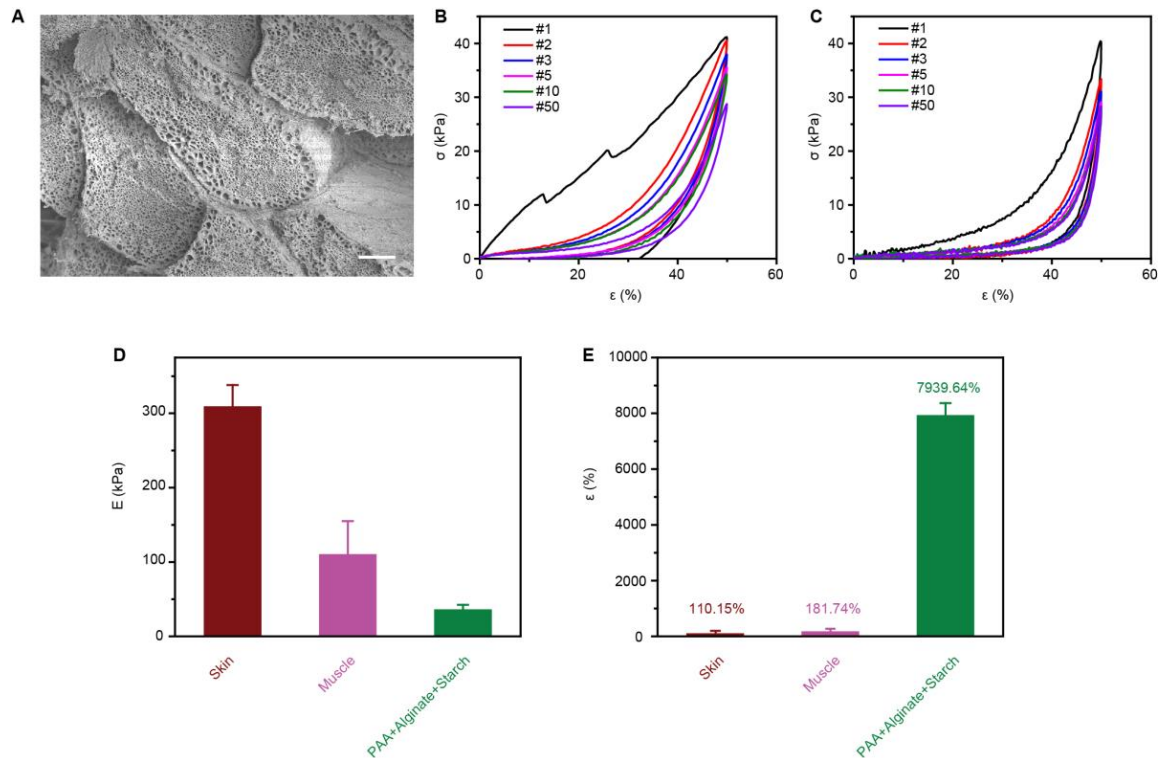

**Figure S1. Structural and mechanical properties of biological tissues. A**, A cross-sectional SEM image of a mouse muscle tissue, showing dense cellular packing in ECM. Scale bar, 10 μm. **B,** A mouse muscle tissue was loaded for 50 cycles under stretch to $\varepsilon = 50\%$. The first cyclic loading-unloading loop exhibits the largest hysteresis. All subsequent cycles until the 50$^{th}$ one had much smaller and overlapping hysteresis. **C**, A mouse skin tissue was loaded for 50 cycles under stretch to $\varepsilon = 50\%$. The first cyclic loading-unloading loop exhibits the largest hysteresis. All subsequent cycles until the 50$^{th}$ one had much smaller and overlapping hysteresis. **D** and **E**, Young's moduli (*E*) and Fractured strains ($\varepsilon_{max}$) from tensile tests for mouse muscle (*n=5*) and starch hybrid gel (*n=5*).**D** and **E**, Young's moduli (*E*) and Fractured strains ($\varepsilon_{max}$) from tensile tests for mouse muscle (*n=5*) and starch hybrid gel (*n=5*).

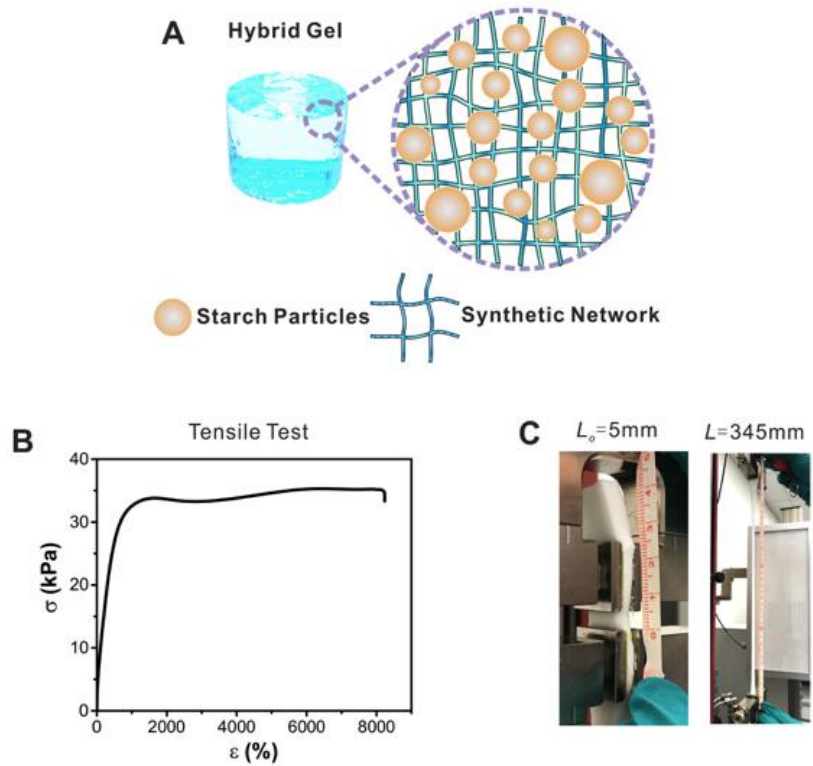

**Figure S2. Structural and mechanical properties of hydrogel composite. A**, A schematic showing a synthetic polymeric network while keeping the starch granular intact, the as-prepared tissue-like hybrid gel. **B**, Stress-strain curve of a 43 wt% starch hybrid gel demonstrating superior stretchability. **C**, Photographs showing the stretching process of a hybrid gel from 5 mm to 345 mm.

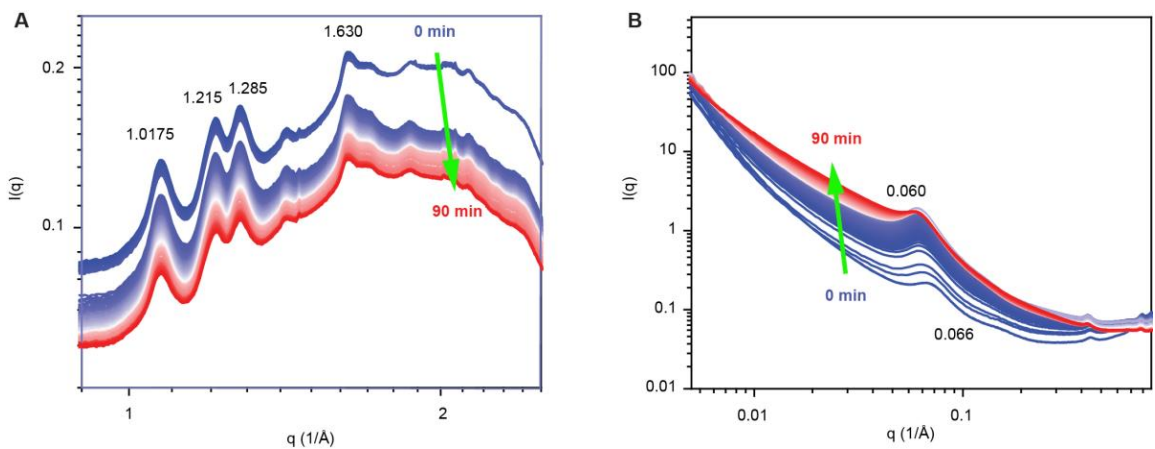

**Figure S3. Time resolved SWAXS during the hydration process of starch granules in the hydrogel composite. A**, WAXS patterns recorded during the hydration process at the different time from 0 min to 90 min. **B**, SAXS patterns recorded from the starch hybrid gel showing the lamellar growth during the hydration process from 0 min to 90 min.

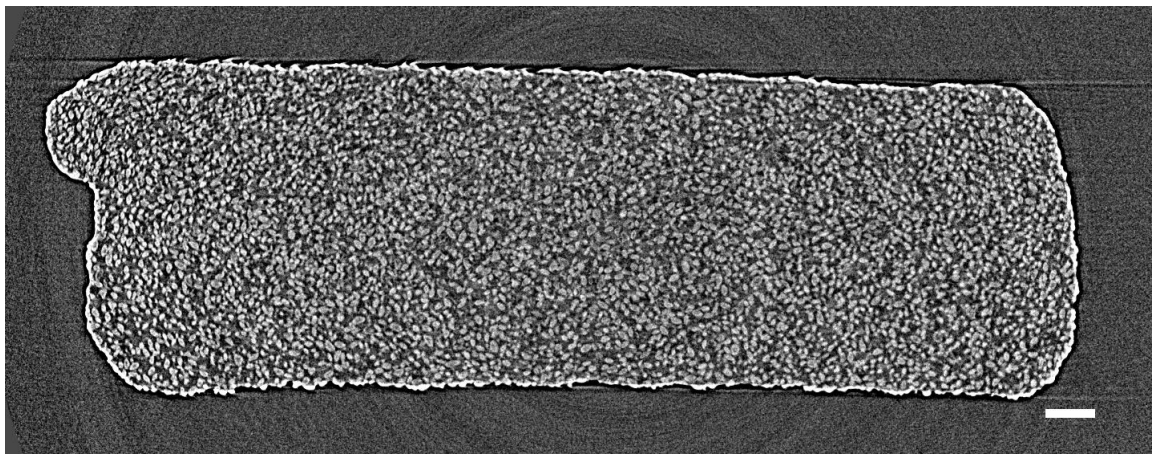

**Figure S4. A representative tomographic reconstructed slice image of a 35 wt% starch hydrogel composite at $\varepsilon$ = 500% stationary state.** Scale bar, 100μm.

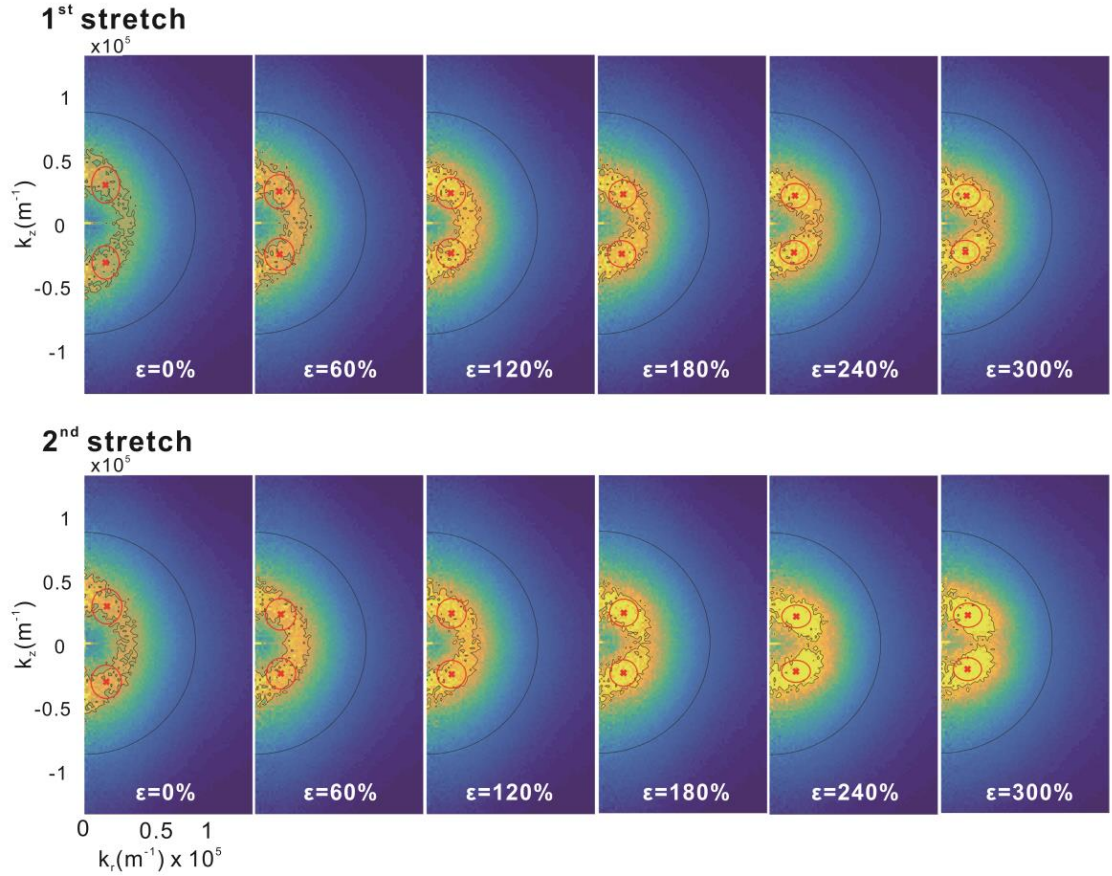

**Figure S5. 2D FFT spectra of a 20 wt% starch hydrogel composite at different strains during the 1st and 2nd stretching processes.** Colormaps show the Fourier amplitude, where yellow represents a high amplitude, and blue represents a low amplitude. The black semi-circles indicate one average particle diameter (11 µm in real space, $9.1\times10^4$ m$^{-1}$ in reciprocal space). Red crosses label the weighted mean positions of the bright regions with the red ellipses corresponding to standard deviations. In both stretching processes, the spectra showed elongations in the $z$-direction.

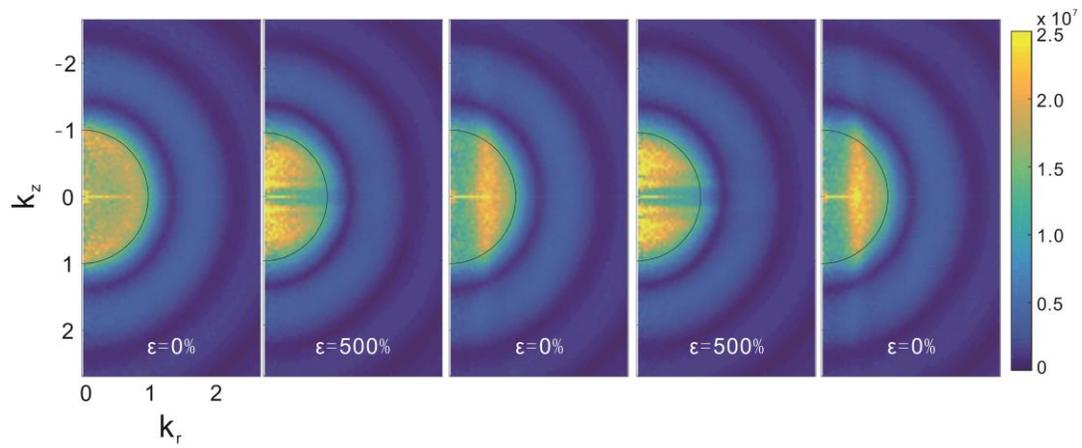

**Figure S6. Numerically simulated Fourier spectra of a hydrogel composite being stretched in the *z*-direction twice.** The particle diameter was set to 1 such that $k = 1$ corresponds to the size of one particle, as labeled by the solid black line. The 1-3 images from left are simulated spectra during the first stretch cycle (*e.g.*, relaxed state→strained state→relaxed state). The 3-5 images from left are the spectra during the second stretch cycle.

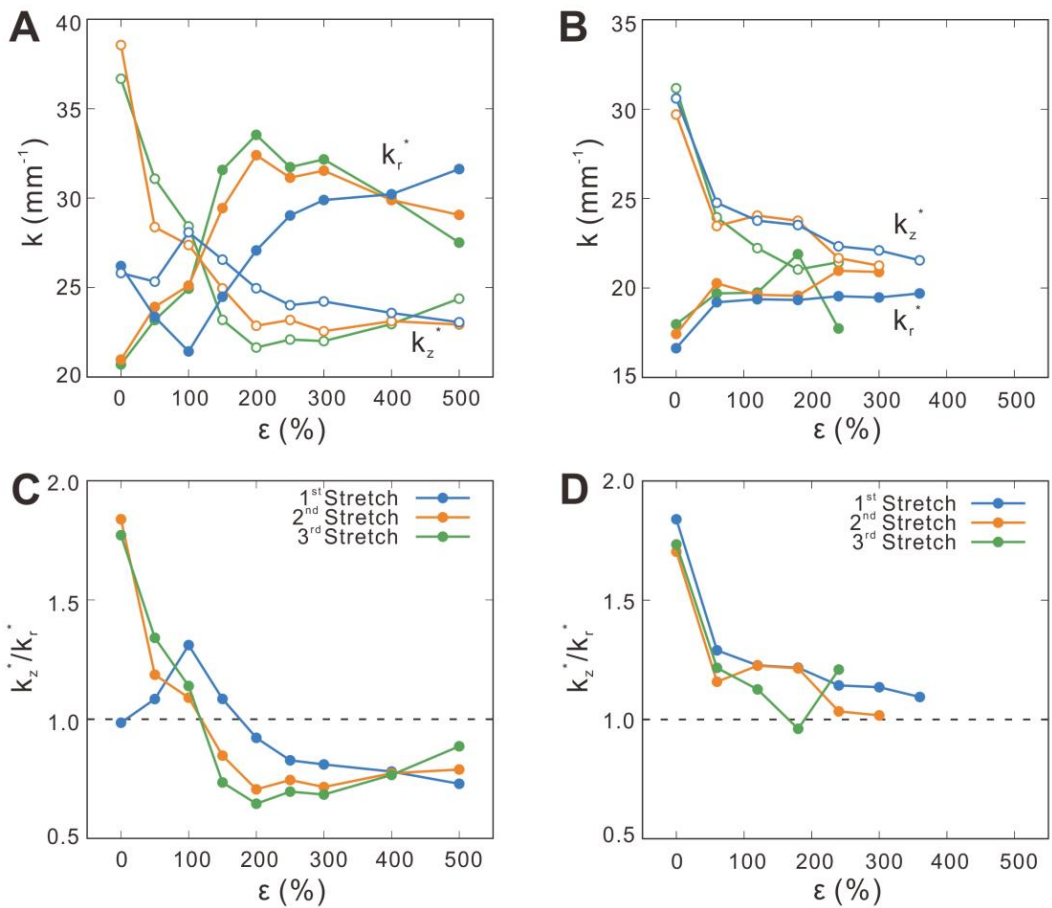

**Figure S7. $k_z^*$ and $k_r^*$ as a function of strain from gels with different starch contents. A** and **B**, Absolute values of $k_z^*$ (open circles) and $k_r^*$ (solid circles) at different $\varepsilon$ in the first (blue), second (orange), and third (green) stretching cycles of 35 wt% (**A**) and 20 wt% (**B**) starch hybrid gels. **C** and **D**, The ratio between $k_z^*/k_r^*$ as a function of strain for samples at 35 wt% (**C**) and 20 wt% (**D**) starch hybrid gels. These results suggest that the packing fraction of the starch granules in the hybrid gel is critical for the observed particle re-organization under strain.

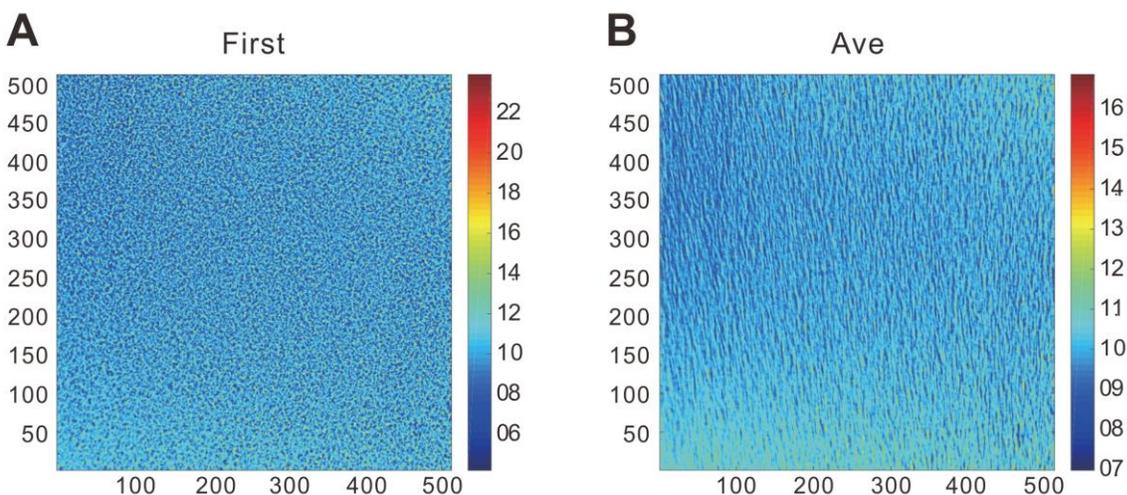

**Figure S8. Coherent X-ray scattering images to directly observe the relaxation of a hydrogel composite. A**, Speckle pattern within a single frame acquired using the near-field coherent X-ray scattering during the stretching and relaxation of a starch hybrid gel. The pattern was due to the phase contrast at the boundaries of the starch grains. Each pixel on the CCD image correspond to 2.56 µm on the sample. Use of the transmission geometry achieved reasonable visibility of speckles collected from an X-ray beam with a longitudinal coherence length 100 times smaller than the monochromatic X-ray beams typically used in coherent X-ray scattering experiments. In addition, the near-field speckles directly relate the motion of the speckles to the motion of the starch particles. **B**, Time-averaged pattern from over 700 continuous frames at a frame rate of 500 Hz. The smearing of the speckles and the occurrence of streaks along the vertical direction within the entire field of view clearly indicates continuous translational motion along the vertical direction, which is consistent with the direction of the pulling and the consequential relaxation.

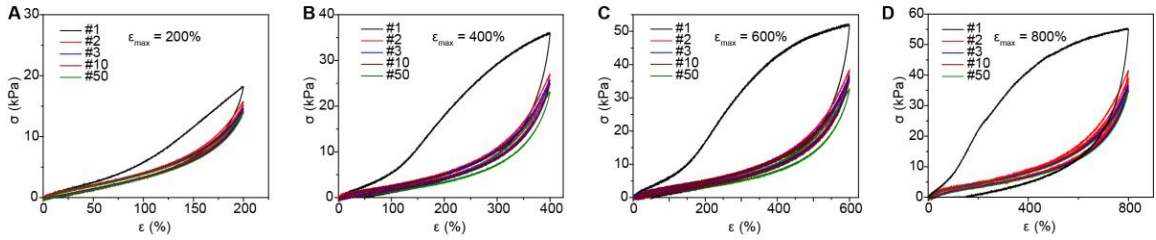

**Figure S9. Fatigue test of starch hybrid gel. A-D**, The starch hybrid gel was loaded for 50 cycles under stretch at $\varepsilon$ =200%, 400%, 600%, 800%. The first cyclic loading-unloading loop exhibits the largest hysteresis. All subsequent cycles until the 49$^{th}$ one had much smaller and overlapping hysteresis. During the measurement, we also marked the clamping positions and removed the sample from the grips every five cycles. After reattaching the sample for subsequent tensile tests, we did not observe the recovery of the largest hysteresis collected during the first loop. The residual strains in the hybrid gels (35 wt% starch in PAA+Alg) are much smaller and the stress-strain behaviors (after the first cycle) are more reproducible, these features are critical for the mechanically encoded memory effect in **Fig. 5**.

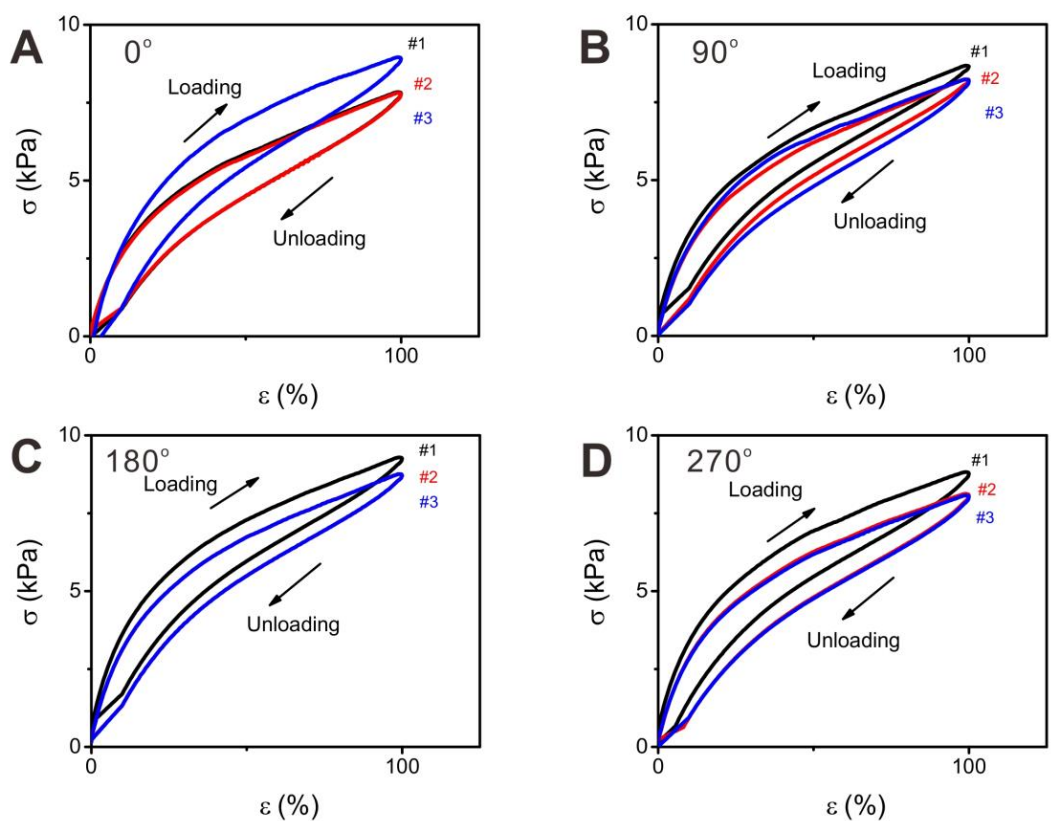

**Figure S10. Energy dissipation from angular loading of 35 wt% starch hydrogel composite up to $\varepsilon = 100\%$.** The square sample ($20 \times 20 \times 5$ mm$^3$) was cyclically loaded three times up to $\varepsilon = 100\%$ along the 0° (**A**), 90° (**B**), 180° (**C**), 270° (**D**) directions. There is negligible hysteresis of each cyclic loading and no apparent difference in energy dissipation of the angular loading at $\varepsilon = 100\%$.

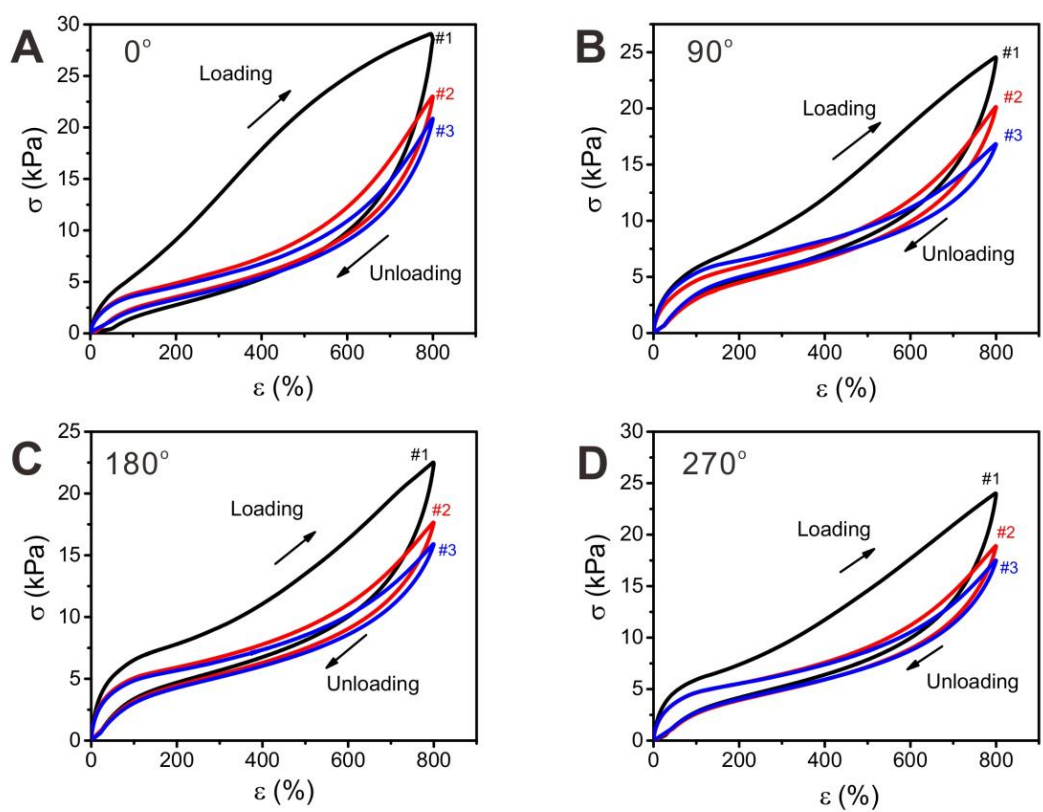

**Figure S11. Energy dissipation from angular loading of 35 wt% starch hydrogel composite up to $\varepsilon$ = 800%. A-D**, The square sample (20 × 20 × 5 mm$^3$) was cyclically loaded three times up to $\varepsilon$ = 800% along the 0° (**A**), 90° (**B**), 180° (**C**), 270° (**D**) directions. The first loading-unloading loop exhibits the most pronounced hysteresis. In sharp contrast, the second and third loading-unloading loops have much smaller areas. When the square sample was loaded in the orthogonal direction (90°), the pronounced hysteresis loop reappeared at the first loading cycle and disappeared at the second and third loops. Further cyclic loading/unloading of the same sample in the other orthogonal directions (180°, 270°) display similar behavior as that observed at 90°, suggesting a reconfigurability of the gel.

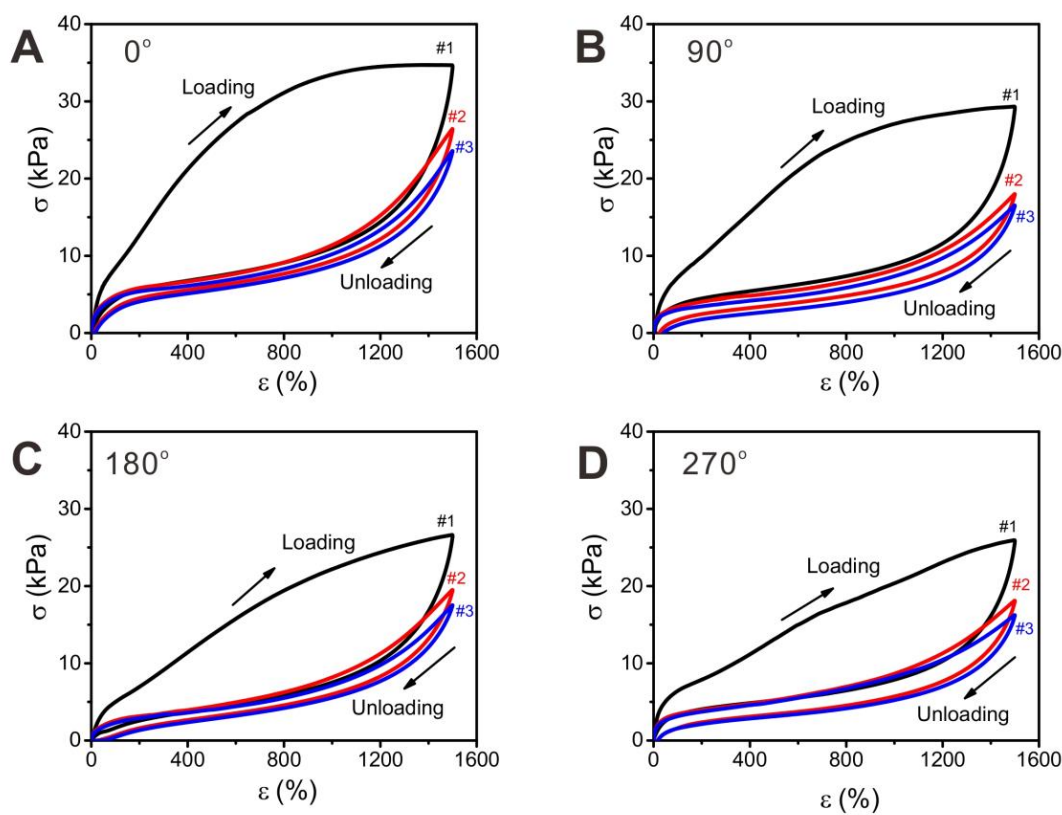

**Figure S12. Energy dissipation from angular loading of 35 wt% starch hydrogel composite up to $\varepsilon = 1500\%$. A-D**, The square sample ($20 \times 20 \times 5$ mm$^3$) was cyclically loaded three times up to $\varepsilon = 1500\%$ along the 0° (**A**), 90° (**B**), 180° (**C**), 270° (**D**) directions. The first loading-unloading loop exhibits the most pronounced hysteresis. In sharp contrast, the second and third loading-unloading loops have much smaller areas. When the square sample was loaded in the orthogonal direction (90°), the pronounced hysteresis loop reappeared at the first loading cycle and disappeared at the second and third loops. Further cyclic loading/unloading of the same sample in the other orthogonal directions (180°, 270°) suggest the reconfigurability of the gel.

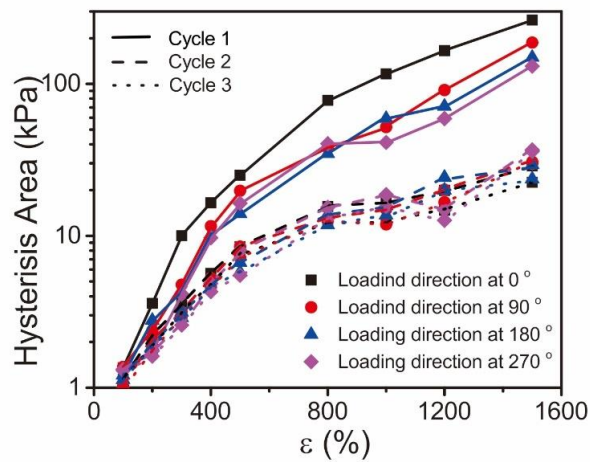

**Figure S13. Strain-dependent energy dissipation during the angular loading of 35 wt% starch hydrogel composite.** In general, the hysteresis area increases with larger strains. At a lower strain (*e.g.,* 100%), the difference between hysteresis areas from different cycles is small, no matter which direction or index of the loading. At higher strains, the gel always dissipates the largest energy in the first cycle along each direction while subsequent cycles showed greatly reduced hysteresis areas. The results suggest the realization (*i.e.*, to turn on a memory device) and performance (*i.e.*, the efficacy of WRITE and OVER-WRITE operations) of the memory device shown in **Fig. 4** can be controlled by the amplitude and orientation of the applied strain.

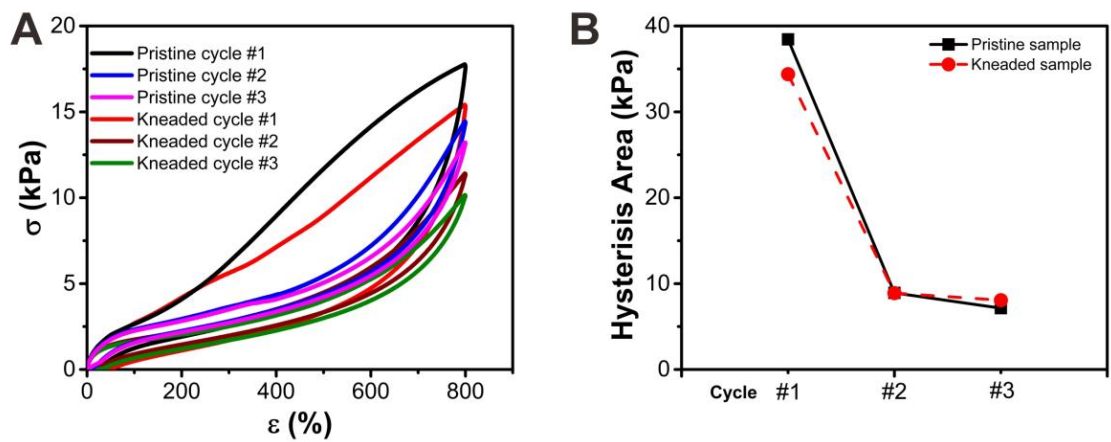

**Figure S14. Energy dissipation of a hydrogel composite after kneading. A**, Cyclic loading and unloading loops of the same starch hybrid gel from its pristine and kneaded states. The pristine sample was first loaded for three cycles and removed from the grips. Subsequent kneading of the sample was able to recover the initial hysteresis area to a large extent. **B**, Quantitative analysis of the hysteresis areas for the pristine and kneaded conditions demonstrates that rejuvenation of starch packing structures by kneading can effectively restore the original energy dissipation of the gel.

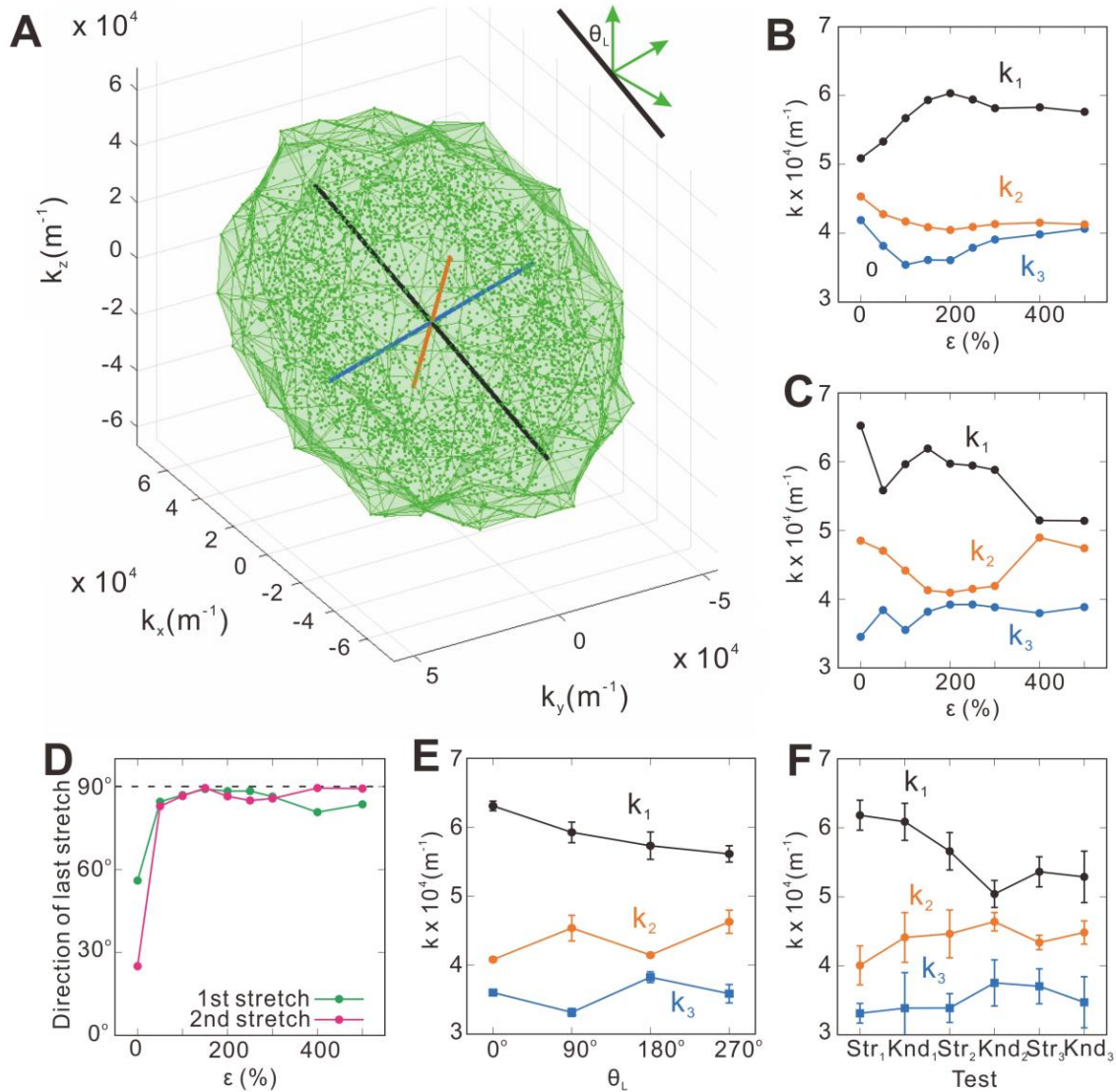

**Figure S15. 3D FFT analysis from X-ray micro-CT in the *k*-space. A**, Positions of top 5000 bright points in *k*-space obtained from 3D FFT of the X-ray images from starch hybrid gel. The data were fitted with an ellipsoid, with its three principal axes labeled: Black, longest axis; blue: shortest axis; orange: intermediate axis. **B** and **C**, Lengths of the three principal axes as a function of strain during the 1st (**B**) and the 2nd (**C**) stretching processes of the starch hybrid gel. During the 1st stretching process, the shape of the ellipsoid became more elongated. In the 2nd stretching process, it started from an ellipsoidal shape with one very long axis ($k_1$) and became slightly more isotropic at larger strains. These all agree with the results obtained from FFT. **D**, Angle between the longest principal axis and the *z*-direction at different strains during the 1st and 2nd stretching processes. The initial angle at $\varepsilon = 0$ before the 1st stretching process is relatively random because the shape of the ellipsoid in k space is more isotropic. Then it quickly rose to 90°. At the beginning of the 2nd stretching process, the long axis is perpendicular to the $z$-direction, which is shown in the main text as well. **E**, Lengths of the axes during stretching along different directions. The overall trend of $k_2$ and $k_3$ are both roughly constant, while $k_1$ showed a slight decrease. The zigzag pattern shows that to higher order, the $\gamma$ and $\gamma'$ states are not exactly the same. Two regions from each of the two X-ray tomography data were analyzed to plot the mean and standard deviation values. **F**, Lengths of the axes in three stretching-kneading cycles. Five regions from each of the two X-ray tomography data were analyzed to plot the mean and standard deviation values.

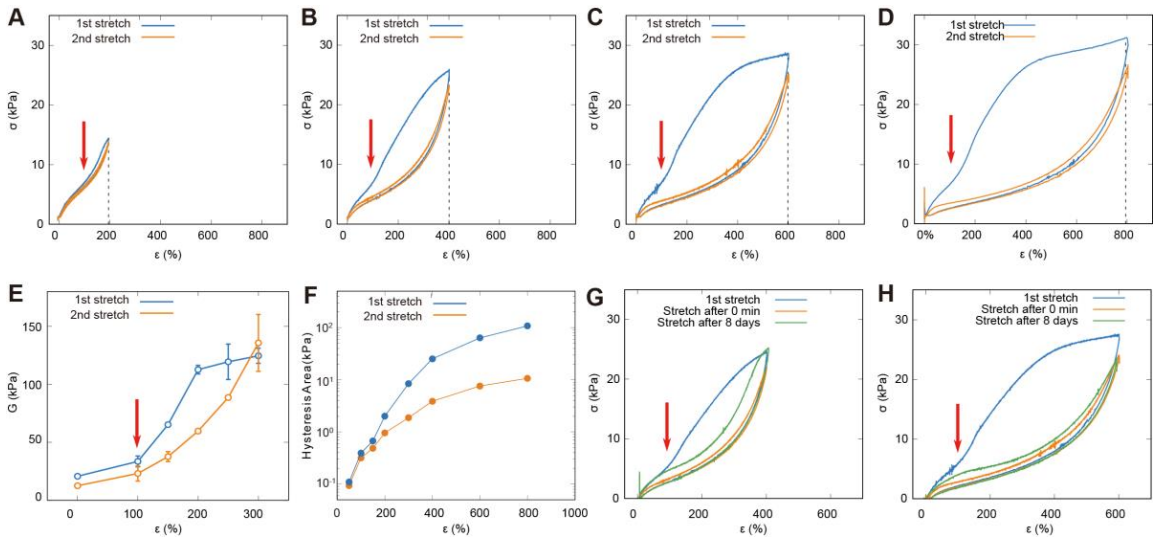

**Figure S16. Transition point for mechanical properties at strain $\varepsilon = 100\%$. A-D**, Stress-strain curves from tensile tests of different starch hybrid gels being stretched to different maximum strains. Each sample was stretched to the same maximum strain twice. The slope increased consistently at $\varepsilon \approx 100\%$ during the first stretching process, *i.e.*, strain stiffening was exhibited. **E**, Shear modulus $G$ as measured by ultrasound shear wave elastography shows similar strain stiffening after $\varepsilon \approx 100\%$, where $G$ increased dramatically from 30 kPa ($\varepsilon = 100\%$) to 110 kPa ($\varepsilon = 200\%$). Three measurements were performed at each strain for statistics. Open circles represent the mean values of the shear modulus. Error bars denote standard deviations. **F**, Hysteresis area enclosed by the loading/unloading curves at different strains in the first (blue) and second (orange) stretches. The upper and lower branches start to bifurcate significantly after $\varepsilon \approx 150\%$. **G** and **H**, Long-term relaxation test of the starch hybrid gel shows minor recovery of the entire hysteresis by dynamic hydrogen bonding via thermal motion. Nevertheless, at smaller strains ($\varepsilon < 100\%$), stress-strain curves almost overlap between the 1$^{st}$ stretch (blue) and the 3$^{rd}$ stretch after 8 days (green). Specifically, newly prepared gels were first stretched to 400% and 600% strain and returned to the original length. A second cycle was performed immediately to evaluate the instantaneous recovery of the hysteresis. After the second test, samples were stored for 8 days in an enclosed environment to avoid dehydration. The results overall suggest that strain-dependent mesoscale starch particle re-arrangement (which cannot be recovered by thermal motion 8 days after the first stretching process, **G** and **H**) is directly related to the observed strain-stiffening behaviors (**A** to **E**), as both sets of data highlight the same transition point of $\varepsilon \approx 100\%$.

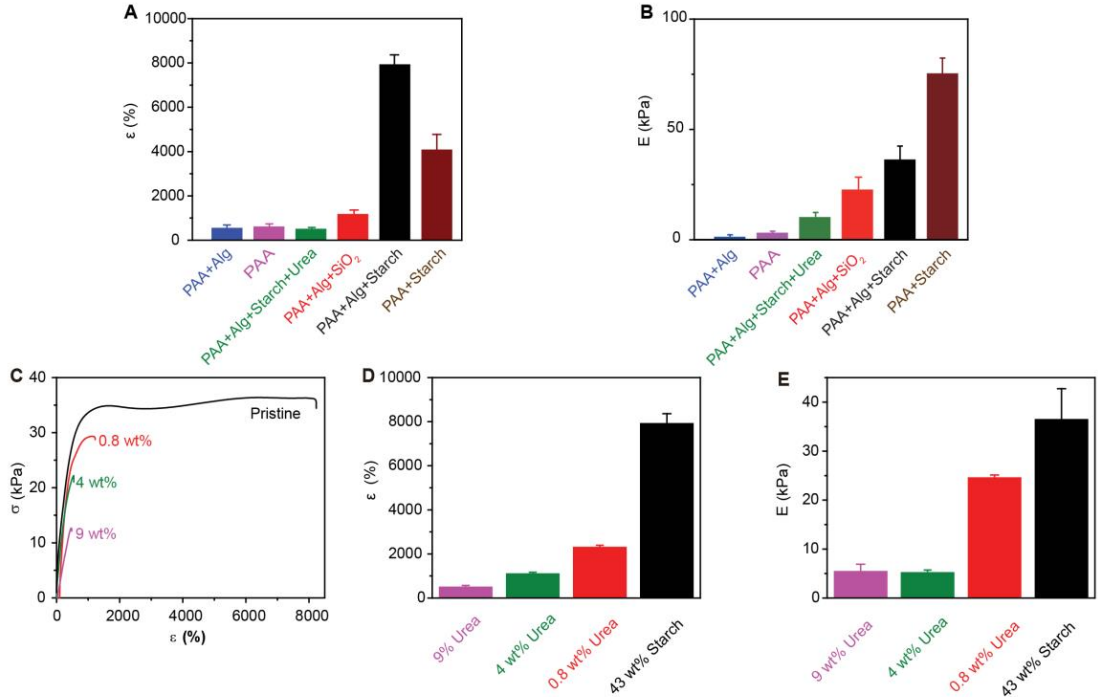

**Figure S17. Quantitative comparison of fractured strains and Young's moduli from different samples. A** and **B**, Fractured strains ($\varepsilon_{max}$) and Young's moduli ($E$) from tensile tests for PAA+Alg ($n=4$), PAA ($n=5$), PAA+Alg+starch+urea ($n=3$), PAA+Alg+ SiO$_2$ ($n=4$), 43 wt% PAA+Alg+starch ($n=6$), and PAA+starch ($n=5$) gels. Mean values plus and minus standard deviations are plotted. **C,** Stress-strain curves showing the gradual degradation of stretchability for starch hybrid gel after adding increasing weight percent of urea. **D** and **E**, Fractured strains ($\varepsilon_{max}$) and Young's moduli ($E$) of 43 wt% starch hybrid gel mixed with 0 wt%, 0.8 wt%, 4 wt%, and 9 wt% urea. The $\varepsilon_{max}$ decrease from ~ 7939% to ~ 2323% when added 0.8% urea to the 43% starch hybrid gel, eventually the $\varepsilon_{max}$ decreases from ~ 7939% to the ~ 519% when the concentration of urea increased to 9%. Moreover, Young's moduli ($E$) measured by the tensile test also present decline when the level of urea gradually increases in the system, the value of Young's modulus eventually reduces from 36.56 ± 6.18 kPa to 5.30 ± 0.42 kPa. The results highlight the critical contribution of hydrogen bonds in the material mechanical properties. Six replicates for the 43 wt% starch hybrid gel and three replicates for each urea-incorporated sample were performed for statistics. Mean values plus and minus standard deviations are plotted.

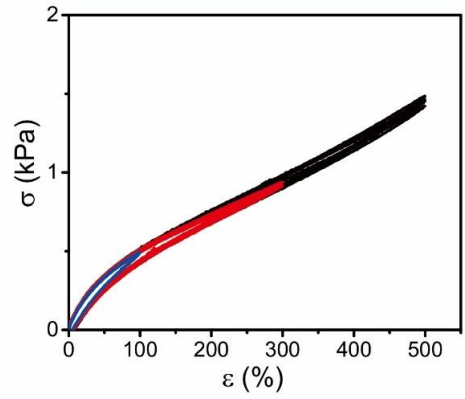

**Figure S18. Mechanical properties of pure PAA-Alg hydrogel.** No apparent energy dissipation of the pure PAA+Alg hydrogel can be observed at various strains. The results highlight the role of starch granules in the observed mechanical behaviors and device demonstration.

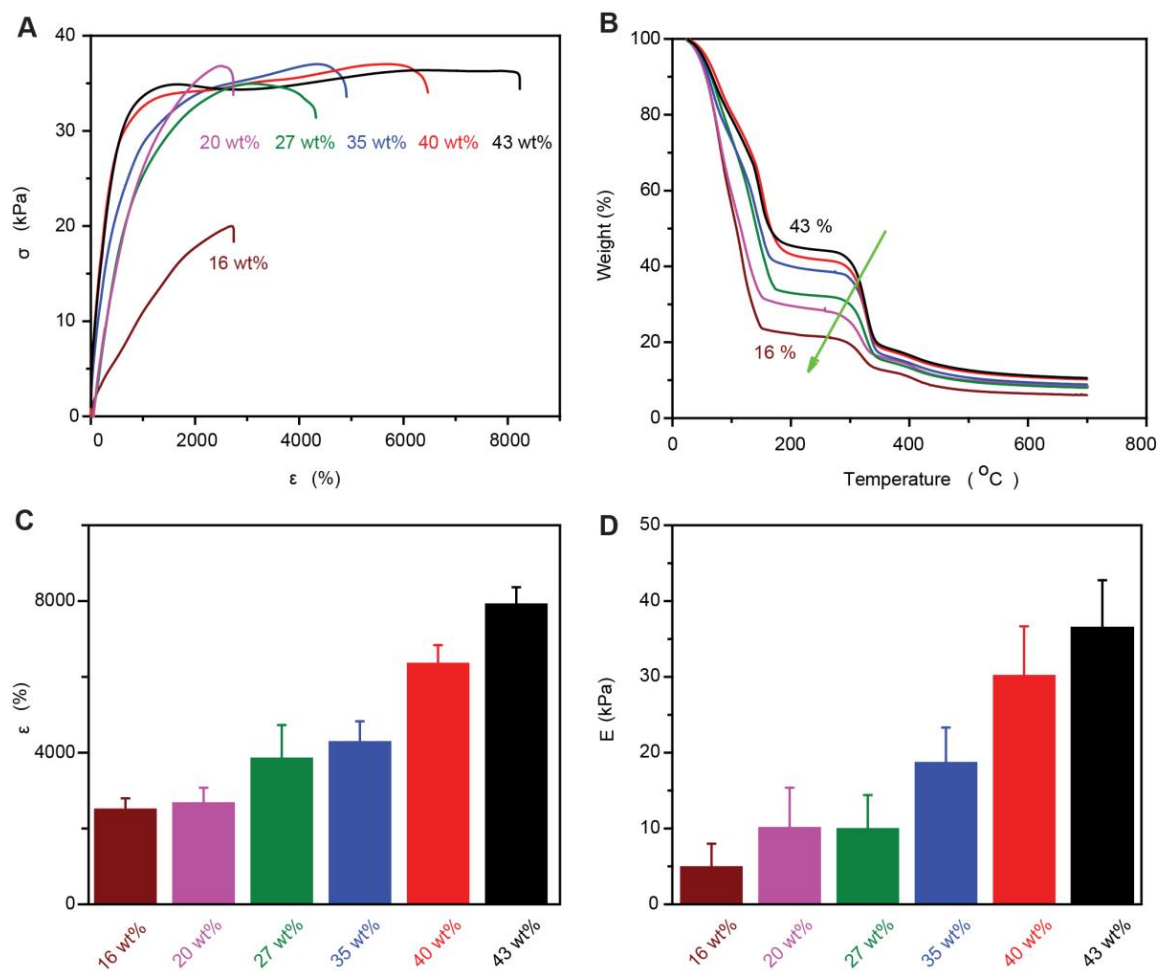

**Figure S19. Mechanical properties of hydrogel composites as a function of starch packing fraction. A,** Stress-strain curves of hydrogels with different starch contents, i.e., 16 wt%, 20 wt%, 27 wt%, 35 wt%, 40 wt%, and 43 wt%. **B**, The thermogravimetric analysis (TGA) results of starch hybrid gels showing the exact water content in each composite. **C, D.** Fracture strains ($\varepsilon_{max}$) and Young's moduli ($E$) quantified by the tensile tests of different starch hydrogels. Gels with higher starch content (*i.e.*, packing fractions > 43 wt%) could not be investigated due to the difficulty of uniformly suspending starch in water. Five replicates for each sample with starch contents from 16 wt% to 40 wt% and six replicates for the 43 wt% starch hybrid gel were performed for statistics. Mean values plus and minus standard deviations are plotted.

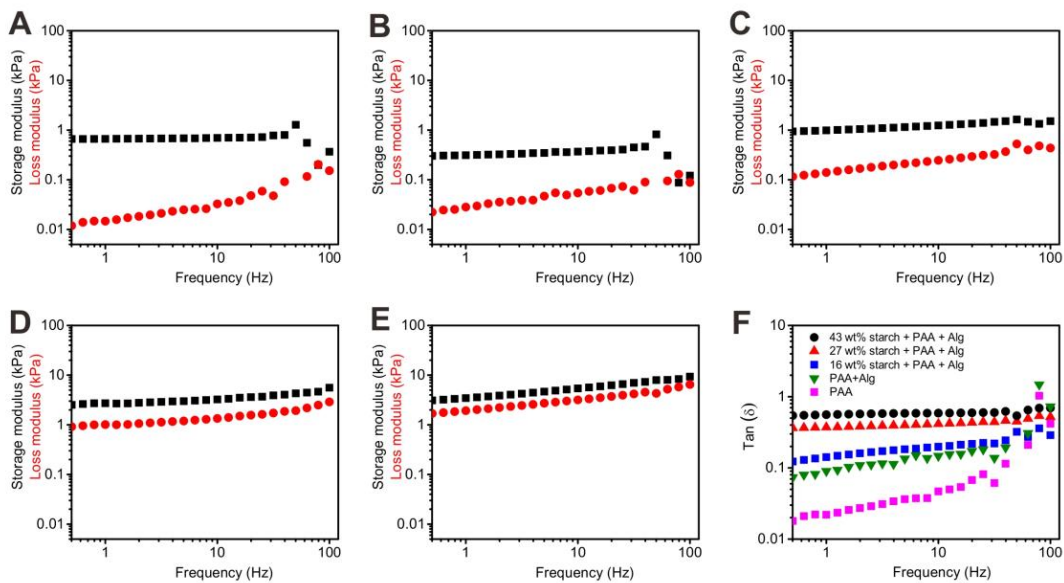

**Figure S20. Rheological properties of different gels.** Storage (black) and loss (red) modulus values of PAA (**A**), PAA + Alg (**B**), 16 wt% starch in PAA+Alg (**C**), 27 wt% starch in PAA+Alg (**D**), and 43 wt% starch in PAA+Alg (**E**). The ratio between loss modulus and storage modulus in (**F**). With the incorporation of Alg in the PAA gel (**B** versus **A**), the loss modulus increased while the storage modulus decreased, indicating more fluid-like features in the PAA+Alg system. With a higher content of starch granules (**C-F**), the ratio between loss and storage moduli increased further, suggesting a trend towards viscoelastic behavior. The compositions for each gel can be found in **Table S1** and **Table S2**.

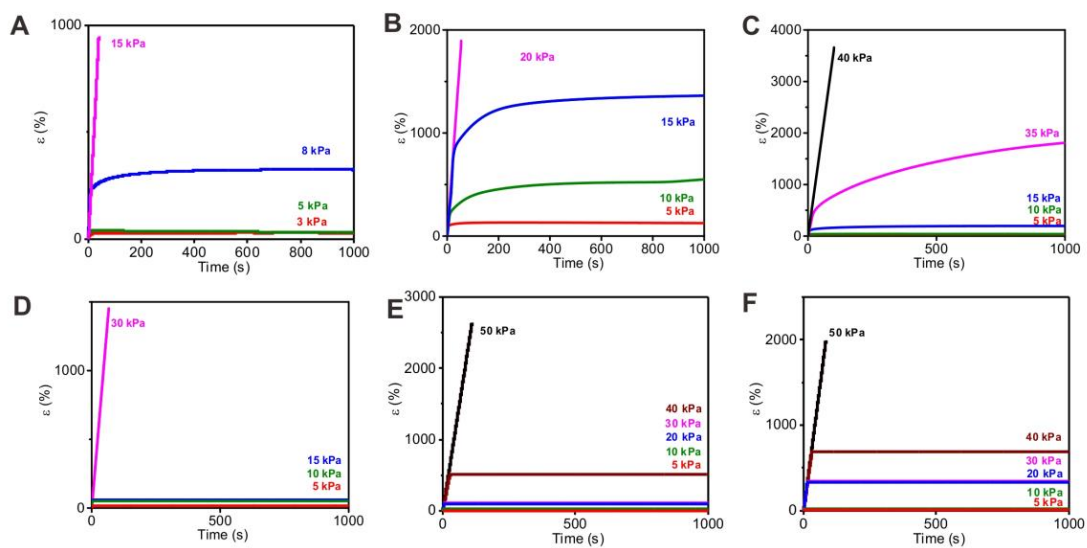

**Figure S21**. **Tensile creep curves of different gels at room temperature.** The sample was loaded with constant tensile stresses as labeled in individual panels following 16 wt% starch in PAA+Alg (**A**), 27 wt% starch in PAA+Alg (**B**), 43 wt% starch in PAA+Alg (**C**), 16 wt% starch in PAA (**D**), 27 wt% starch in PAA (**E**), and 43 wt% starch in PAA (**F**). Gels with starch granules in PAA+Alg (**A**-**C**) exhibits more fluid-like when compared to gels with starch in PAA only (**D**-**E**) under the same starch content. With the larger amount of starch, a higher stress is required to induce a significant flow of the gel. The compositions for each gel can be found in **Table S1** and **Table S2**.

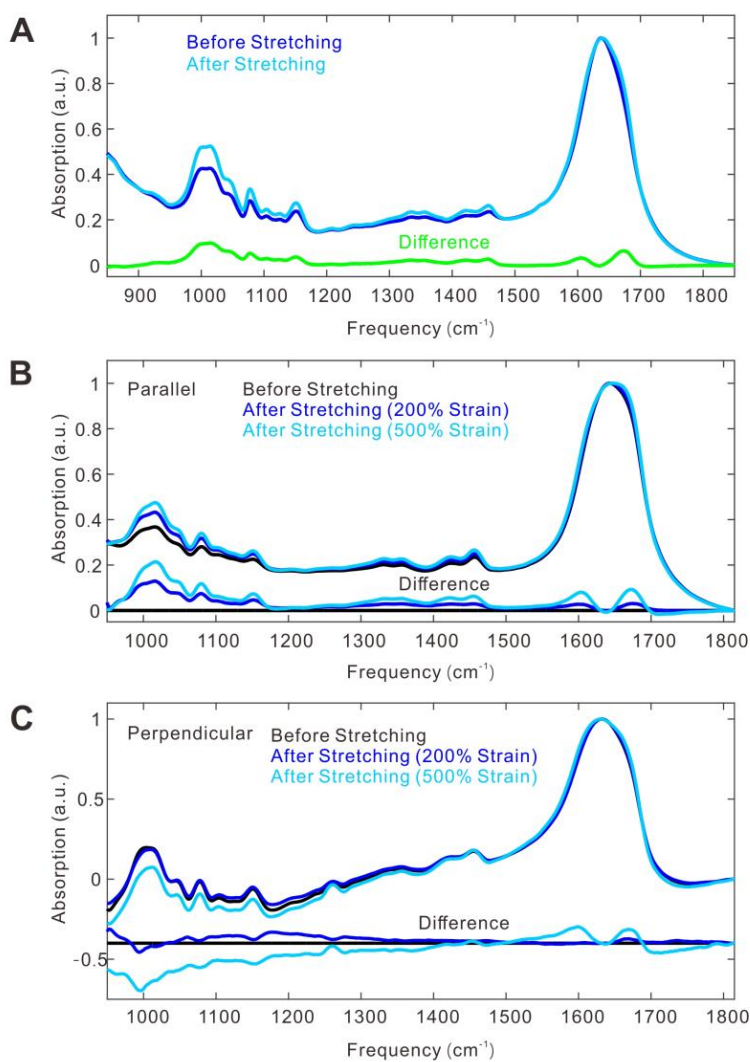

**Figure S22. Enhancement of hydrogen bond networks in the hydrogel composite upon stretching. A**, ATR-FTIR spectra of a starch hybrid gel before (blue) and after (cyan) stretching to 500% strain, with their difference spectrum (green) plotted at the bottom. **B** and **C**, Polarized ATR-FTIR spectra of a starch hybrid gel with parallel (**B**) and perpendicular polarization (**C**). Black, unstretched; blue, 200% strain; cyan, 500% strain. Difference spectra are shown underneath with corresponding colors.

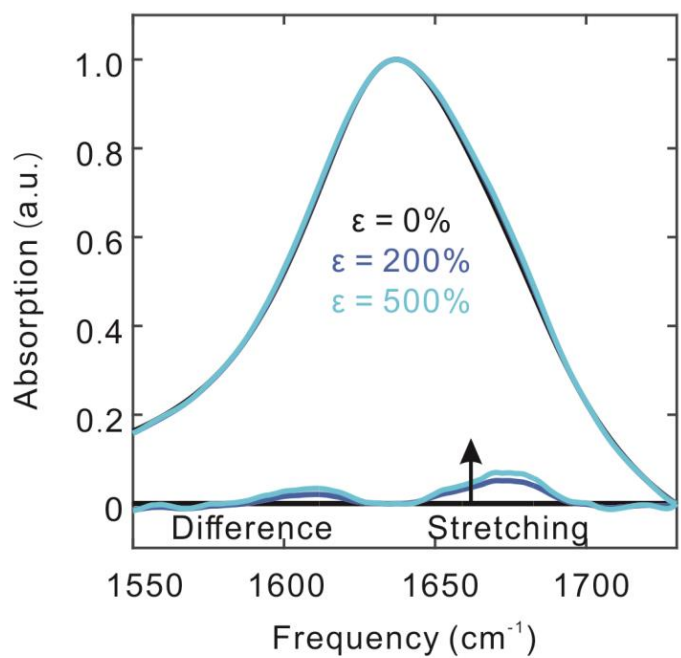

**Figure S23. ATR-FTIR spectra of the same SiO$_2$-filled gel at different strain levels.** Difference spectra with the scale factor of 3, are plotted at the bottom with respect to the unstrained sample (black trace). Blue and cyan traces represent data from $\varepsilon = 200\%$ and $\varepsilon = 500\%$, respectively. When compared with the data shown in **Fig. 5B**, the spectra suggest that the starch particle/PAA interface can better enhance the hydrogen bond network.

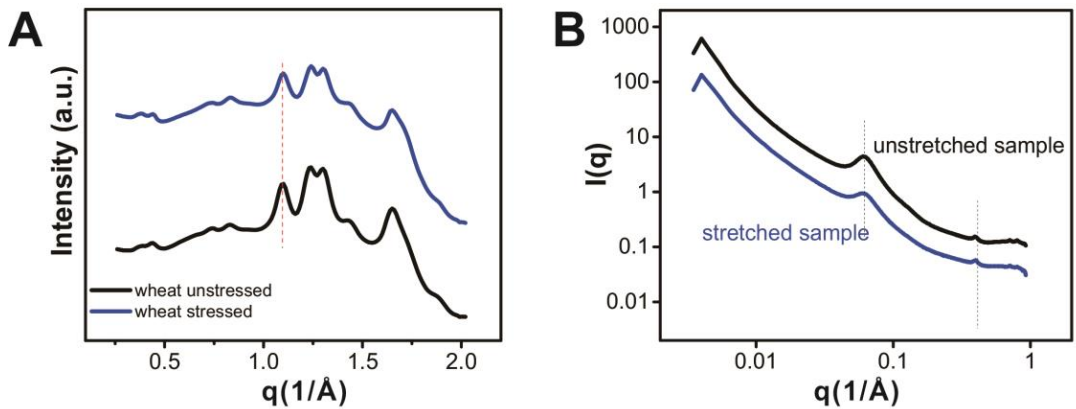

**Figure S24. WAXS and SAXS of a starch hybrid gel confirms the stability of the inherent starch structures before and after stretching to 1000%. A**, Extended WAXS shows that all characteristic diffraction peaks of the starch double helical structure remain unchanged upon mechanical stretching. **B**, SAXS patterns recorded from the starch hybrid gel shows that the lamellar structure of starch remains unchanged upon mechanical stretching. These results suggest that the observed mechanical properties (**Figs. 2, 3, 4,** and **6**) are not relevant to the inner structure of starch.

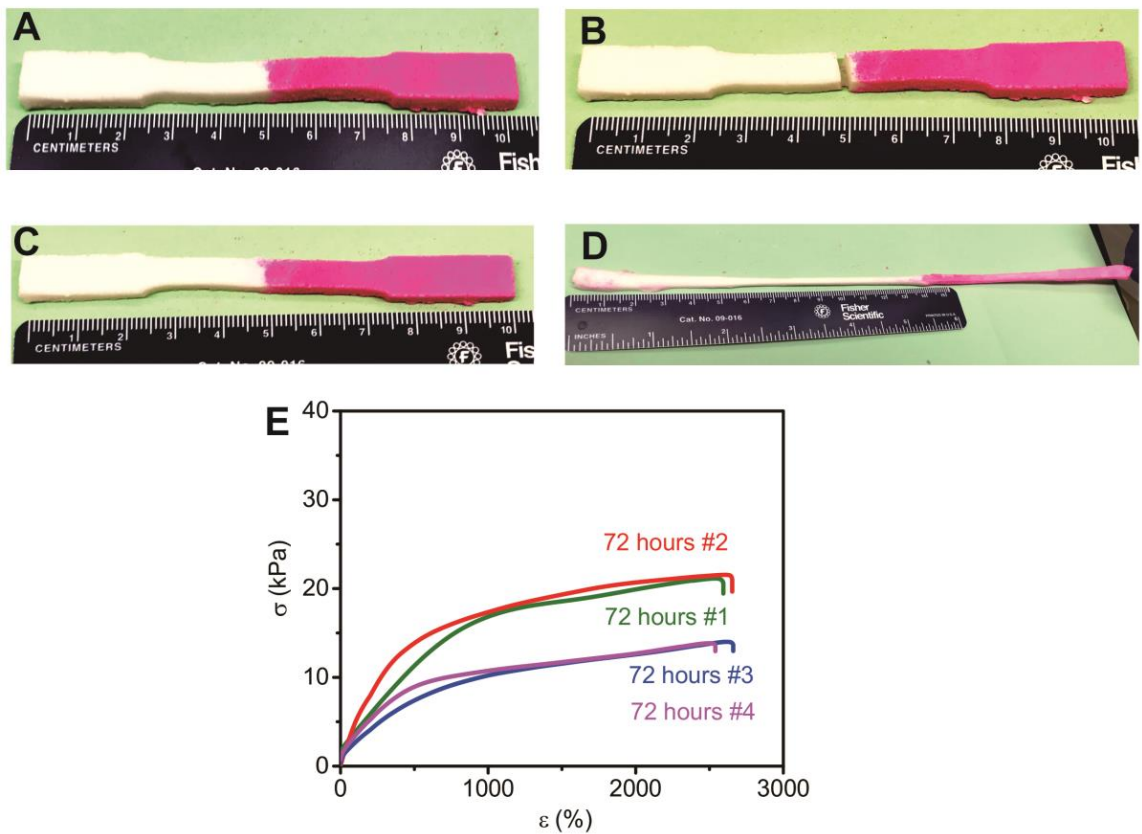

**Figure S25. Self-healing of a hybrid gel (27 wt% starch in PAA+Alg).** Photographs (**A-D**) of the pristine starch hybrid gel (**A**), the same gel with a line cut right next to the white/red boundary (**B**), the same gel after being self-healed for 72 hours at ambient condition (**C**), and the as-healed sample being stretched (**D**). **E**, Strain-stress curves of four different gels after healing, suggesting a reproducible self-healing behavior. The self-healing demonstration highlights the dynamic hydrogen bonding in the hybrid gel system.

# 3. Supplementary Tables
Table S1. Detailed composition of different starch hybrid gels.

## Hybrid Hydrogel Network 1

|  | Wheat Starch (wt%) | Acrylamide (wt%) | Alginate (wt%) | Water (wt%) | APS+MBAA (wt%) |
|---|---|---|---|---|---|
| Network 1 | 42.78 | 3.42 | 0.21 | 53.48 | 0.11 |
| Network 2 | 38.85 | 3.66 | 0.23 | 57.14 | 0.12 |
| Network 3 | 34.63 | 3.94 | 0.24 | 61.07 | 0.12 |
| Network 4 | 27.21 | 4.35 | 0.27 | 68.02 | 0.15 |
| Network 5 | 20.22 | 6.48 | 0.42 | 72.71 | 0.17 |
| Network 6 | 15.75 | 5.04 | 0.31 | 78.73 | 0.17 |

**Table S2. Composition of different types of gels.**

### Hybrid Hydrogel Network 2

| | Silica* (wt%) | Wheat Starch (wt%) | Acrylamide (wt%) | Alginate (wt%) | Water (wt%) | APS+MBAA (wt%) |
|---|---|---|---|---|---|---|
| Category 1 | | 42.78 | 3.42 | 0.21 | 53.48 | 0.11 |
| Category 2 | 38.76 | | 4.96 | 0.31 | 55.81 | 0.17 |
| Category 3 | | | 5.98 | 0.37 | 93.44 | 0.20 |
| Category 4 | | | 6.00 | | 93.79 | 0.20 |
| Category 5 | | 42.87 | 3.43 | | 53.58 | 0.12 |
| Category 6 | | 27.3 | 4.40 | | 68.20 | 0.10 |
| Category 7 | | 15.80 | 5.00 | | 79.10 | 0.10 |

Note: Silica* is diols-functionalized silica.

**Table S3. A summary of the multi-scale characterizations utilized to demonstrate the "Dynamic and programmable cellular-scale granules enable tissue-like materials".**

| Length scale | Experiments | Key results | Figure index |
|---|---|---|---|
| Nano- and Meso- | X-ray Micro-CT imaging. | Starch granules are tightly packed in the gel matrix. | Fig. 1B |
| Nano- | X-ray Nano-CT imaging. | Individual starch granules have irregular shapes, and their surfaces have nanoscale roughness. | Fig. 1C |
| Meso- | Cryo-SEM imaging of starch hybrid gel. | The nanoscale hydrogel matrix seamlessly connected to the mesoscale granular particles in starch hybrid gel. | Fig. 1D |
| Meso- | 2D FFT spectra of 3D reconstructed X-ray micro CT imaging during the 1st and 2nd stretching processes. | Applied strain induced reorganization of starch granules in receptacle space during the 1st and 2nd stretching processes | Fig. 2, B-D, SI figs. S4, 5, 7 |
| Meso- | Coherent X-ray scattering image of a starch hybrid gel | The relaxation dynamics of a starch hybrid gel | Fig. 3, A-D, SI fig. S8 |
| Macro- | Cyclic tensile tests to study the energy dissipation for the different stretches at the different strain. Fatigue test of starch hybrid gel | The energy dissipation capability of the starch hybrid gel is dependent on its history of external stress. | Fig. 4, A-C |
| Macro- | Starch hybrid gels coated on a robotic hand | The artificial skin with heterogeneous compliance after *in situ* mechanical training, promising for robotic skin applications | Fig. 4, E |
| Macro- | Cyclic tensile tests at different orientations to study the energy dissipation. | The energy dissipation capability of the starch hybrid gel shows the reconfigurable effect during orthogonal mechanical loading | Fig. 5, A-D, SI figs. S10-13 |
| Meso- | 3D FFT spectra of 3D reconstructed X-ray micro-CT in the *k*-space during the orthogonal stretching processes. | Revealed the reorganization and memory effect of starch granules in receptacle space during the orthogonal stretching processes. | Fig. 5E, SI fig. S15 |
| Macro- | Photographs of the setup for ultrasound shear wave elastography | Shear wave measurement can be done during in-situ stretching. | Fig. 6A |
| Macro- | Shear wave imaging with high-speed ultrasound. | To observe the shear wave image by space-time plots of the position and precisely calculate shear modulus by $G = \rho c_s^2$. Meanwhile, the real time monitoring the variation of shear modulus shows the strain-stiffness at $\varepsilon \approx 100\%$. | Fig. 6, C-D, |
| Macro- | Tensile tests of hydrogels with different material composition. | Starch granules could significantly enhance the mechanical properties. | Fig. 7A, SI fig. S17 |
| Macro- | Tensile tests of hydrogels with Urea additive | Hydrogen bond exists in the starch hybrid gel and critically affected the mechanical properties. | Fig. 7A, SI fig. S17 |
| Molecular | ATR-FTIR of the starch hybrid gels during stretching, recorded in $H_2O$. | Applied strain enhanced hydrogen bonding in the starch hybrid gel during the stretching process. | Fig. 7B, SI fig. S22 |
| Macro- | Tensile tests at different time points during self-healing | Hydrogen bond exists in the starch hybrid gel and critically affected the mechanical properties. | Fig. 7C, SI fig. S25 |

| Macro- | X-ray microCT imaging during self-healing | The cut starch hybrid gel can recover its morphology upon self-healing. | Fig. 7D |
| --- | --- | --- | --- |
| Macro- | Ball drop and vertical impact and tests showing the significantly enhanced impact absorption of the hydrogel composite versus the control PAA+Alg gel. | Starch granules could significantly enhance the mechanical properties. | Fig. 7E |
| Macro- | Vetical impact tests | Starch granules could significantly enhance the impact absorption and reduce fluctuations. | Fig. 7, F-I |
| Macro- | Photographs of robotic hands partially coated with starch hybrid gel as artificial skin. | Photographs perfectly presents human machine interfeace. | Fig. 8C |
| Nano- and Meso | Cross-sectional SEM image of a mouse muscle tissue. | The cross-sectional SEM image of a mouse muscle tissue reveals the structure of myofibrils and myosin at the multi-length scale. | SI fig. S1A |
| Macro- | Cyclic tensile tests to study the energy dissipation of a mouse tissue and skin for the different stretches at $\varepsilon \approx 50\%$. | To study energy dissipation of the mouse muscle and skin, figure out the similarity of our starch hybrid gel. | SI fig. S1B |
| Macro- | Tensile tests of mouse skin and muscle. | To study the maximize fracture strain and the effective Young's modulus. | SI fig. S1, D, E |
| Meso- and Meso | Structural and mechanical properties of starch hybrid gel. | To study the basic mechanical properties and microstructure of starch hybrid gel. | SI fig. S2, A-D |
| Nano- | The in situ SWAXS recoding for the starch hybrid gel during the hydration process. | To study the dynamic hydration process of starch hybrid gel. | SI fig. S3 |
| Meso- | Numerical simulation of starch hybrid gel during stretching. | Simulated FFT spectra shows a positive Poisson's ratio for the starch hybrid gel. | SI fig. S6 |
| Macro- | Cyclic tensile tests to study the energy dissipation for the different stretches at the different strain. Fatigue test of starch hybrid gel. | The energy dissipation capability of the starch hybrid gel is dependent on its history of external stress. | SI fig. S9 |
| Macro- | Cyclic tensile tests of a starch hybrid gel after kneading to study the energy dissipation. | Kneading can effectively restore the original energy dissipation of the starch hybrid gel. | SI fig. S14 |
| Macro- | Summary the common transition point of mechanical test and ultrasound at $\varepsilon \approx 100\%$. | In the summarized plot, we found that the transition measured by tensile and ultrasound shows the same common point at $\varepsilon \approx 100\%$. The long-term relaxation loading/unloading curve also reveal the partial recovery. | SI fig. S16 |
| Macro- | Cyclic tensile tests to study the energy dissipation of PAA+ Alg hydrogel. | No apparent energy dissipation of the pure PAA+Alg hydrogel can be observed at various strains. | SI fig. S18 |
| Macro- | The tensile test of starch hybrid gel with different packing fraction. | The packing fraction of starch in the starch hybrid gel will significantly affect the mechanical properties. | SI fig. S19, A, C, D |
| Macro- | The thermalgravimetric analysis. | The weight component could be precisely estimate. | SI fig. S19B |
| Macro- | Rheology tests of hydrogels with different material composition. | With a higher content of starch granules, the ratio between loss and storage moduli increased further, suggesting a trend towards viscoelastic behavior. | SI fig. S20 |
| Macro- | Tensile creep tests of different hydrogels. | Gels with starch granules in PAA+Alg exhibits more fluid-like when compared to gels with starch in PAA only under the same starch content. With the larger amount of starch, a higher stress is required to induce a significant flow of the gel. | SI fig. S21 |
| Molecular | Polarized ATR-FTIR of the starch hybrid gels during stretching, recorded in $H_2O$. | Enhancement of hydrogen bond 3D networks in the starch hybrid gel upon stretching. | SI fig. S22 |
| Molecular | ATR-FTIR of the $SiO_2$ hybrid gels during stretching, recorded in $H_2O$. | Applied strain slightly enhanced hydrogen bonding in the $SiO_2$ hybrid gel. | SI fig. S23 |
| Nano- | WAXS of a starch hybrid gel during stretching. | The nanoscale structure in the starch particle of the hybrid gel does not change upon stretching. | SI fig. S24 |
| Nano- | SWAXS of a starch hybrid gel during stretching. | The lamellar structure in the starch particle of the hybrid gel does not change upon stretching. | SI fig. S24 |
| Macro- | Photographs of a starch hybrid gel to show its self-healing behavior. | The cut starch hybrid gel can recover its stretchability upon self-healing. | SI fig. S25 |

## 4. Supplementary Videos

**Movie S1:** Bending of the robotic fingers at the joint positions during the first several cycles locally deformed the hydrogels.

**Movie S2:** Ultrasound shear wave elastography data of a hydrogel composite before stretching.

**Movie S3:** Ultrasound shear wave elastography data of a hydrogel composite being stretched to $\varepsilon = 300\%$.

**Movie S4:** The hydrogel composite film underwent an extreme deformation, when a steel ball dropped on a suspended film, it still maintained its integrity and hold the ball steadily.

**Movie S5:** The film made of PAA-Alg immediately broke upon falling of the ball.

**Movie S6:** A vertical compression test of a hydrogel composite (35 wt% starch) showed negligible lateral fluctuation during the impact process.

**Movie S7:** A vertical compression test of a hydrogel composite (16 wt% starch) showed more obvious fluctuation compared to the 35 wt% composite during the impact process.

**Movie S8:** A vertical compression test of a PAA hydrogel showed significant lateral fluctuation compared to the 35 wt% composite during the impact process.